# Multi-Tier Adaptive Memory Programming and Cluster- and Job-based Relocation for Distributed On-demand Crowdshipping

Tanvir Ahamed[1], Bo Zou.

Department of Civil, Materials, and Environmental Engineering, University of Illinois at Chicago

**Abstract:** With rapid e-commerce growth, on-demand urban delivery is having a high time especially for food, grocery, and retail, often requiring delivery in a very short amount of time after an order is placed. This imposes significant financial and operational challenges for traditional vehicle-based delivery methods. Crowdshipping, which employs ordinary people with a low pay rate and limited time availability, has emerged as an attractive alternative. This paper proposes a multi-tier adaptive memory programming (M-TAMP) to tackle on-demand assignment of requests to crowdsourcees with spatially distributed request origins and destination and crowdsourcee starting points. M-TAMP starts with multiple initial solutions constructed based on different plausible contemplations in assigning requests to crowdsourcees, and organizes solution search through waves, phases, and steps, imitating both ocean waves and human memory functioning while seeking the best solution. The assignment is further enforced by proactively relocating idle crowdsourcees, for which a computationally efficient cluster- and job-based strategy is devised. Numerical experiments demonstrate the superiority of M-TAMP over a number of existing methods, and that relocation can greatly improve the efficiency of crowdsourcee-request assignment.

[1] Corresponding author. Email: tanvir.ahamed@hotmail.com

# 1  Introduction

E-commerce has experienced unprecedented growth in the recent years. In the US alone, e-commerce sales achieved $513 billion in 2018, or 9.6% in total retail sales and a 14.3% increase just from the previous year (US census, 2019). The amount of online sales is projected to continue and reach $669 billion by 2021 (Statista, 2018) which accounts for 13.7 percent of all retail sales in US (Statista, 2017). The rapid development of e-commerce along with the fact that the majority of e-commerce sales occurs in cities and increasingly on-demand suggests considerable increase in delivery traffic. Relying still largely on vehicle asset-based business models, it means more dispatching of delivery vehicles, which results in many negative consequences including increased VMTs, traffic congestion, illegal street parking, greenhouse gas emissions, air pollution, and wear-and-tear of road infrastructure, which are increasingly at odds with the need and trend of developing livable and sustainable urban environment.

The increasing on-demand nature of city logistics, especially among deliveries of food, grocery, and retail, has imposed considerable pressure on delivery service providers (DSP) to control logistics cost while meeting the customer demand. These deliveries are often of small size, and from distributed origins (restaurants, grocery stores, and retail shops rather than the DSP's warehouse) to distributed destinations (customers in residence or workplaces). It has become increasingly popular that customers want to receive the delivery after an order is placed within half to a few hours. Given the vast volume of such orders, it would be operationally and financially impractical for DSPs to constantly dispatch self-owned vehicles for pickup and delivery of such services. This is because, first of all, the number of self-owned vehicles is unlikely to be sufficient to meet the demand. Even if a DSP has enough vehicles, it would be too expensive to do so.

As a result of the pressure, innovations are being pushed to meet the growing on-demand, small-size, and distributed demand with short, guaranteed delivery time. Among the innovations, crowdshipping has emerged as an attractive, low-cost alternative. In crowdshipping, a DSP solicits ordinary people, termed crowdsourcees who have available time and may walk, bike, or drive to perform delivery to earn income. Employing crowdsourcees as *ad-hoc* couriers often requires a lower payment rate than full-time employees, thus bringing DSPs significant cost advantage. As crowdsourcees are either non-motorized or using smaller personal cars, the negative consequences from DSP-owned truck-/van-based delivery can be largely mitigated. In the business practice, crowdshipping is rapidly developing with companies like UberEats, Grubhub, DoorDash, Postmates,

Deliv, Piggy Baggy, Amazon Flex, and Dada-Jing Dong Dao Jia reshaping the dinning, e-grocery, and online retail businesses.

This paper focuses on developing methodologies for dynamic assignment of requests to crowdsourcees, and for proactively relocating idle crowdsourcees to mitigate spatial supply-demand imbalance in the crowdshipping system. As the delivery environment considered in the paper is on-demand, assignment and relocation decisions need to be made in real time while ensuring the quality of the decisions. For the assignment problem, while it bears some relevance with the dynamic pickup and delivery problem with time windows (PDPTW) in the vehicle routing problem (VRP) literature, the problem does differ in that crowdsourcees have limited and heterogenous amount of available time, and the time between the generation of a request and the latest delivery of the request is very short. A novel population- and memory-based neighborhood search heuristic is developed which yields better results than existing methods for solving PDPTWs. For the relocation problem, given that a real-world system can involve a non-trivial number of idle crowdsourcees, a cluster- and job-based strategy is conceived for relocation decisions in a computationally efficient manner. The performance of joint and periodic assignment and relocation is evaluated in extensive, large-size numerical experiments, with the results compared with those obtained from using the state-of-the-art methods.

Three major contributions are made in the paper. First, a novel multi-start, multi-tier adaptive memory programming (M-TAMP) algorithm is proposed for assigning requests to crowdsourcees in an on-demand environment. The multiple tiers include decomposing the solution search through waves, phases, and steps. By doing so, the neighborhood search for better solutions is made more systematic and effective given the huge number of routing permutations to explore. Strategic oscillation among solutions is embedded, which allows solutions in the memory to be reexamined, solutions that repeatedly appear in the memory to be tabued to avoid cyclic search, and currently not-so-good but promising solutions to be kept in memory for further improvement. We also investigate the construction of initial solutions for the algorithm, which is shown critical to the search for best solutions. All these contribute to the computational superiority of M-TAMP over existing methods. Second, the possibility to improve system responsiveness to new shipping requests and reduce total shipping cost is investigated. Specifically, a clustering-based strategy is proposed that makes relocation decisions in a computationally effective and efficient manner. Third, through comprehensive numerical experiments, a number of managerial insights are offered about crowdshipping system operations to meet on-demand urban deliveries.

The remainder of the paper is structured as follows. Section 2 reviews relevant literature. A formal description of the problem for on-demand request-crowdsourcee assignment and idle crowdsourcee relocation is presented in Section 3. In Section 4, we provide an overall picture of the M-TAMP algorithm for the assignment problem. This is followed by detailed description of the elements in M-TAMP in Section 5. Section 6 presents the cluster- and job-based relocation strategy with its mathematical formulation. Extensive numerical experiments are conducted in Section 7. The findings of the paper are summarized and directions for future research are suggested in Section 8.

## 2 Literature review

Our review of relevant literature is organized in three subsections. First, we review existing research on planning and operation of crowdshipping systems. Then, methods for solving dynamic PDPTW, a problem closely resembling our assignment problem, are reviewed. The last subsection focuses on research on transportation resource relocation, which is relevant to relocating idle crowdsourcees.

### 2.1 Crowdshipping system design and operation

As a result of the rapid development in industry practice, crowdshipping has garnered much research attention in recent years. Relevant literature considers three aspects: supply, demand, and operation and management (Le et al., 2019). As our paper concerns the third aspect, the review in this subsection mainly focuses on different operational schemes and strategies that have been proposed.

One branch of crowdshipping research works consider crowdsourcees to perform the last leg of the urban delivery, while propose traditional vehicles to coordinate with the crowdsourcee routes. In the similar direction, a number of operational concepts have appeared in the literature. Kafle et al. (2017) proposes a two-echelon crowdshipping-enabled system which allows crowdsourcees to perform the first/last leg of pickup/delivery while relaying parcels with trucks at intermediate locations. A mixed integer program (MIP) along with tailored heuristics is proposed to minimize system overall cost. Wang et al. (2016) formulate a pop station-based crowdshipping problem in which a DSP only sends parcels to a limited number of pop stations in an urban area, while crowdsourcees take care of delivery from the pop stations to customers. Focusing on the crowdsourcee part, a minimum cost network flow model is formulated to minimize crowdsourcing expense while performing the last-leg deliveries. Macrina et al. (2020) consider that crowdsourcees may pick up parcels from a central depot or an intermediate depot to which parcels are delivered by classic vehicles. Customers can be served either by a classic vehicle or a crowdsourcee. Each crowdsourcee serves at most one customer. To solve this problem, an MIP and a metaheuristic are developed. The use of intermediate transfer locations is further considered by Sampaio et al. (2018) who propose a heuristic for solving multi-depot pickup and delivery problems with time windows and transfers, and by Dötterl et al. (2020) who use an agent-based approach to allow for parcel transfer between crowdsourcees.

Another branch of crowdshipping research considers not only dedicated crowdsourcees but also people "on-the-way" (introduced as occasional drivers), who are willing to make a single delivery for a small amount of compensation using their own vehicles. The static version of this problem is formulated as a vehicle routing problem with occasional drivers, solved by a multi-start heuristic (Archetti et al., 2016). The dynamic version is considered in Dayarian and Savelsbergh (2017) solved by two rolling horizon dispatching approaches: a myopic one that considers only the state of the system, and one that also incorporates probabilistic information about future online order and in-store customer arrivals. A rolling horizon framework with an exact solution approach is proposed by Arslan et al. (2019) which look into dynamic pickups and deliveries using "on-the-way" crowdsourcees. In-store shoppers are also considered as possible crowdsourcees to perform delivery (Gdowska et al., 2018).

Recently, an analytic model is developed by Yildiz and Savelsbergh (2019) to investigate the service and capacity planning problem in a crowdshipping environment, which considers the use of both crowdsourced and company-provided capacity to ensure service quality. The study seeks to answer many fundamental questions such as the relationship between service area and profit, and between delivery offer acceptance probability and profit, and the benefits of integrating service of multiple restaurants. Recognizing that using friends/acquaintances of the shipping request recipients makes delivery more reliable, Devari et al. (2017) conduct a scenario-based analysis to explore the benefits of retail store pickups that rely on friends/acquaintances on their routine trips to stores/work/home. Similarly, Akeb et al. (2018) examine another type of crowd logistics in which neighbors of a customer collect and deliver the parcels when the customer is away from home, thereby reducing failed deliveries.

Despite the proliferation of research in the crowdshipping field, an efficient solution technique for dynamic PDPTW, while proactively relocating idle crowdsourcees simultaneously, has not been considered in the literature as a way to improve the crowdshipping system operations. In addition to this major gap, there seems to be little attention paid to the limited time availability that dedicated crowdsourcees are likely to have while performing crowdshipping. This paper intends to address these gaps.

## 2.2 Methods for solving dynamic PDP

As mentioned in Section 1, the crowdshipping problem considered in this paper can be viewed as an adapted version of dynamic PDPTW, for which the reader may refer to Berbeglia et al. (2010) and Pillac et al. (2013) for reviews. Given the NP-hard nature of PDPTW, it is not surprising that most existing solution methods for dynamic PDPTW are heuristics.

A most widely used heuristic for dynamic PDPTW is tabu search. Two notable works in this field are Mitrović-Minić et al. (2004) and Gendreau et al. (2006). The former considers jointly a short-term objective of minimizing routing length and a long-term objective of maximizing request slack time while inserting future requests into existing routes. The latter employs a neighborhood search structure based on ejection chain. However, neither of the two studies account for future requests. Ferrucci et al. (2013) develop a tabu search algorithm to guide vehicles to request-likely areas before arrival of requests for urgent goods delivery. This work is later extended by Ferrucci and Bock (2014) to include dynamic events (e.g., new request arrival, traffic congestion, vehicle disturbance).

Other than tabu search, Ghiani et al. (2009) employ an anticipatory insertion procedure followed by an anticipatory local search to solve uncapacitated dynamic PDPTW for same-day courier service. Similar to the idea of Bent and Van Hentenryck (2004) to evaluate potential solutions using an objective function that considers future requests through sampling. The study exploits an integrated approach in order to address multiple issues involved in real-time fleet management including assigning requests to vehicles, routing and scheduling vehicles, and relocating idle vehicles. The authors find that the proposed procedure for routing construction outperforms their reactive counterparts that do not use sampling. Besides heuristics, an optimization-based algorithm is proposed by Savelsberg and Sol (1998) to solve real-world dynamic PDPTW.

## 2.3 Resource relocation

To our knowledge, no research exists on crowdsourcee relocation. If we consider idle crowdsourcees as resources, crowdsourcee relocation resembles the problem of relocating empty vehicles which has been investigated in a few different contexts. An example is autonomous mobility systems, in which relocating idle autonomous vehicles can better match vehicle supply with rider demand (Fagnant and Kockelman, 2014; Zhang and Pavone, 2016). The need for vehicle relocation also arises in one-way carsharing. Boyaci et al. (2017) and Nourinejad et al. (2015) show the potential benefit of relocating empty electric vehicles to better accommodate carsharing demand. Sayarshad et al. (2017) propose a dynamic facility location model for general on-demand service systems like taxis, dynamic ridesharing services, and vehicle sharing. An online policy is proposed by the authors to relocate idle vehicles in a demand responsive manner with look-ahead considerations.

Relocating vehicles has been studied, at least implicitly, for freight movement as well. In the PDPTW literature, Van Hemert and La Poutré (2004) introduce the concept of "fruitful regions" to explore the possibility of moving vehicles to fruitful regions to improve system performance. Motivated by overnight mail delivery services, a traveling salesman problem is investigated by Larsen et al. (2004) with three proposed policies for vehicle repositioning based on *a-priori* information. Waiting and relocation strategies are developed in Bent and Van Hentenryck (2007) for online stochastic multiple vehicle routing with time windows in which requests arrive dynamically and the goal is to maximize the number of serviced customers.

## 3 Problem description

As mentioned in Section 1, this paper investigates a specific type of on-demand crowdshipping that the origins and destinations of shipping requests and the starting points of crowdsourcees are all spatially distributed. A DSP constantly receives requests and uses crowdsourcees to pick up and deliver the requests within a short, guaranteed period of time (e.g., two hours). Crowdsourcees also dynamically enter the crowdshipping system. Crowdsourcees have limited available time and will exit the system when the available time is used up. Each crowdsourcee can carry multiple requests as long as their carrying capacity allows. Crowdsourcee routes can be dynamically adjusted as new incoming requests are added at the end or inserted in the middle of the routes. With proper payment as incentives, idle crowdsourcees can move from their current locations to new locations suggested by the DSP. The relocation is to benefit the DSP by reducing its total shipping cost (TSC).

In this paper, we consider that the DSP performs periodical crowdsourcee-request assignment, each time with the latest information about unassigned shipping requests and crowdsourcees, who can be either idle or en route, carrying previously assigned requests. As such, a request can be inserted at the end or middle of an existing crowdsourcee route. Alternatively, a request may be given to an idle crowdsourcee. After a request is generated, it will be considered at the next immediate assignment. However, it is possible that some requests cannot be assigned to any existing crowdsourcees as feasibility and/or pickup distance constraints are violated (see subsection 5.1). In this case, a request may be picked up and delivered by a backup vehicle if urgent, or left for the next assignment.

Each time an assignment is made, decisions on relocating idle crowdsourcees are made right after. Compared to instant assignment whenever a new request or a new crowdsourcee enters the system, periodic treatment can take advantage of accumulated information during the inter-assignment period to make assignment more cost-effective. The period certainly should not be too long given that each request has a delivery time guarantee. The extent to which the inter-assignment period length on system performance will be investigated in future computational experiments.

# 4 M-TAMP algorithm for request-crowdsourcee assignment: Overall picture

## 4.1 Fundamental idea

The proposed M-TAMP algorithm is a multi-start, population- and memory-based neighborhood search heuristic. The algorithm starts from multiple initial solutions constructed based on different plausible contemplations for assigning requests to crowdsourcees. Through neighborhood search, a growing number of alternative solutions are generated and evaluated in an iterative manner, with a portion of them selected to enter an adaptive memory ($AM$) in which solutions are constantly updated and direct further solution search. The selection of solutions to enter/update $AM$ is probabilistic, thus allowing some currently not-so-good but maybe promising solutions to be selected, which may get substantially improved in subsequent search. $AM$ plays the role as a container of good and not-so-good but maybe promising solutions which, like human memory, evolves in the course of the search. Solutions that show repeated appearance will be permanently kept (tabued) in $AM$.

The novelties of the M-TAMP heuristic lie in three parts: 1) structuring the solution search by constructing sequential and interconnected waves. Each wave is associated with removing requests from one route and inserting the removed requests to other routes; 2) structuring each wave by alternately performing vertical and horizontal phases. A vertical phase pertains to adding solutions to $AM$, whereas a horizontal phase reexamines solutions in $AM$; 3) structuring each horizontal phase by conducting removal-replacement of $AM$ solutions in multiple steps. Below we provide more detailed (though still conceptual) elaboration of the three novelties.

For the first novelty, given that a solution entails a number of routes, instead of improving all routes simultaneously, the M-TAMP heuristic focuses on one route at a time, by moving a request from the route of focus to another route. As there exist many possibilities for a move (considering many requests in the route of focus and many positions in other routes to insert the moved request), only a portion of the possibilities will be selected for consideration. Such a decomposition and solution improvement by focusing on one route at a time prevents dealing with a huge number of neighborhood search all at once, which would be computationally expensive and cumbersome, and lead to the algorithm losing focus in search. This is particularly important as memory is involved in the search. Too many solutions all appearing at the same time will just render the probability of solution reoccurrence very unlikely. In addition, the sequential nature of the decomposition, i.e., solutions resulting from focusing on one route will be used for generating new solutions when focus is shifted to the next route, allows the search efforts made earlier to be used in subsequent routing improvement. The idea of route improvement by focusing on one route at a time is akin to ocean waves lapping at the shore one after another, thus. This explains the use of the term "waves".

For the second novelty, a multitude of permutations exists in each wave because many possibilities exist for picking up a request to remove from the route of focus and inserting the removed request in another route. Exploring all possibilities at once will be computationally expensive. In addition, doing so would generate a vast number of solutions which can result in exceeding randomness in selecting solutions to enter $AM$. Instead, M-TAMP organizes the search by performing routing improvement in consecutive, alternating vertical and horizontal phases. A vertical phase adds solutions to $AM$ (thus the length of $AM$ is vertically expanding), whereas a horizontal phase reexamines the solutions in $AM$ by removing each solution and replacing the solution with another solution probabilistically selected from $CL$. In a vertical phase, before selecting a solution to enter $AM$, M-TAMP also explores an improvement possibility by removing/inserting one more request for the solutions selected. Iteratively performing solution addition in vertical phases and removal/replacement in horizontal phases follows the idea of a strategic oscillation (Glover and Laguna, 1997a and 1997b), which allows for reexamining the solutions in $AM$. By doing so, $AM$ gets updated and evolves based on the latest neighborhood search towards better solutions.

For the third novelty, in a horizontal phase the removal-replacement operation involves removing each solution from $AM$, returning the solution to $CL$, and probabilistically selecting a solution from $CL$ and add the solution to the end of $AM$ (i.e., replacement). The purpose is to reexamine the solutions in $AM$: if a solution is inferior (high TSC), then the chance of getting the solution back by probabilistic selection is low. Thus, such solutions are removed from $AM$ and replaced by better solutions. For a good solution (low TSC), getting it back to $AM$ is still likely through probabilistic selection. As new solutions are constantly added to $CL$, the removal-replacement operations updates $AM$ by giving chances for good solutions that appear later in $CL$ to replace earlier solutions in $AM$. To avoid the cycling of removing solutions that have repeatedly entered $AM$, we tabu solutions with sufficient reoccurring frequency in $AM$. In performing the removal-replacement, each horizontal phase is decomposed into multiple steps, each step performing removal-replacement on only a portion of the solutions in $AM$. Doing so helps currently not-so-good but maybe promising solutions to get a chance to be selected from $CL$ to enter $AM$. This is because, by removing a fraction of $AM$ solutions at a time and adding them to $CL$, the influence of new $AM$ solutions on selecting an existing solution in $CL$ is limited. This is desirable as it preserves the chance to enter solutions in $AM$ that are currently not so good but could get improved in subsequent waves. Otherwise, removing $AM$ solutions all at once would result in decreased chance for existing solutions in $CL$ to be selected, especially when the $AM$ solutions have low TSC.

### 4.2 Difference of M-TAMP from existing metaheuristics

It is worth highlighting the difference between the M-TAMP algorithm and some existing metaheuristics: genetic algorithm (GA), simulated annealing (SA), and tabu search (TS), which have been widely used in combinatorial optimization including VRP/PDPTW. GA and SA do not involve a memory. For GA, it starts with an initial population of solutions and then iteratively performs genetic operators including crossover and mutations to produce offspring solutions. For efficient search for a given problem, tuning parameters is critical but non-trivial (Smit and Eiben, 2009). SA relies on a single solution in the search. Mimicking the cooling process of materials, SA requires the cooling rate to be very slow when a number of local minima are scattered in the solution space. This is in contrast with the computational requirement for dynamically solving PDPTW in the on-demand delivery context.

TS exploits memory structures in combination with strategic restrictions and aspiration levels as a means for exploring the search space (Glover et al., 1995). There are different types (e.g., classical TS, reactive TS, threshold-based TS, etc.) in the literature. The literature suggests that the solution quality depends on the value of the tabu tenure, i.e., duration that an attribute tabued to avoid cyclic exploration (Glover, 1995). If the tabu tenure value is set too high, then the search will be confined to a specific region of the search space. If the value is set too low, then the chance of cycling in solution exploration will increase, compromising computational efficiency. Research on this topic is abundant but still needs further development to present a robust and deterministic tabu scheme. In addition, while the concept of tabu-list in TS bears a resemblance to AM, the use and construction mechanism of AM (remove-replace) is very different.

## 4.3 Overall flow of M-TAMP

Following the idea discussed above, this subsection provides further details about the M-TAMP algorithm, graphically shown in Fig. 1 and procedurally exhibited in Algorithm 1 below. Table 1 provides the notations used in this subsection and Section 5.

**Table 1.** Notations used in the M-TAMP algorithm and their definitions

| | |
|---|---|
| **Lists and Sets** | |
| $CL$ | Candidate list of routing solutions |
| $CL'$ | Temporary candidate list, which is a set of routing solutions |
| $AM$ | List of solutions in the adaptive memory |
| $S_{PA}$ | Set of solutions that have entered $AM$ more than a certain number of times |
| **Variables** | |
| $R_0$ | Initial solution |
| $C_{R_0}$ | Shipping cost of solution $R_0$ |
| $G$ | Guaranteed delivery time |
| $T_{R,m}$ | Travel time on a route $m$ of a solution $R$ |
| $T_{R,g}$ | Amount of lateness for delivery with respect to the delivery time guarantee for all the assigned requests in the routes of a solution |
| $T_{R,avl}$ | Amount of time violation with respect to the available time window over all crowdsourcees for a solution |
| $N_{R,cap}$ | Total number of requests that exceed the carrying limits over all crowdsourcees for a solution |
| $\Pr(R)$ | Probability of selecting a candidate solution $R$ |
| $R_1$ | Selected candidate solution from $CL$ to apply forward move |
| $R_1'$ | Selected solution in the vicinity of the selected solution ($R_1$) |
| $R_1''$ | Generated forward move by application of intra-route move on $R_1'$ |
| $R_{best}$ | Best solution after improvement of initial solution $R_0$ using M-TAMP |
| $C_{best}$ | Best solution cost after improvement of initial solution $R_0$ using M-TAMP |
| **Parameters** | |
| $\alpha$ | Parameter that emphasizes or de-emphasizes the cost reduction potential of a solution |
| $\mu_q$ | Number of solutions in $AM$ during phase $q$ |
| $h_q$ | Number of horizontal steps required during phase $q$ |
| $d_{i,q}$ | Number of remove/replace move required from/to $AM$ during a horizontal step $i$ of a phase $q$ |
| $\delta$ | Parameter that facilitates the calculation of number of phase reqd. ($q^*$) |
| $\eta_q$ | Separation between successive $\mu_q$ values |

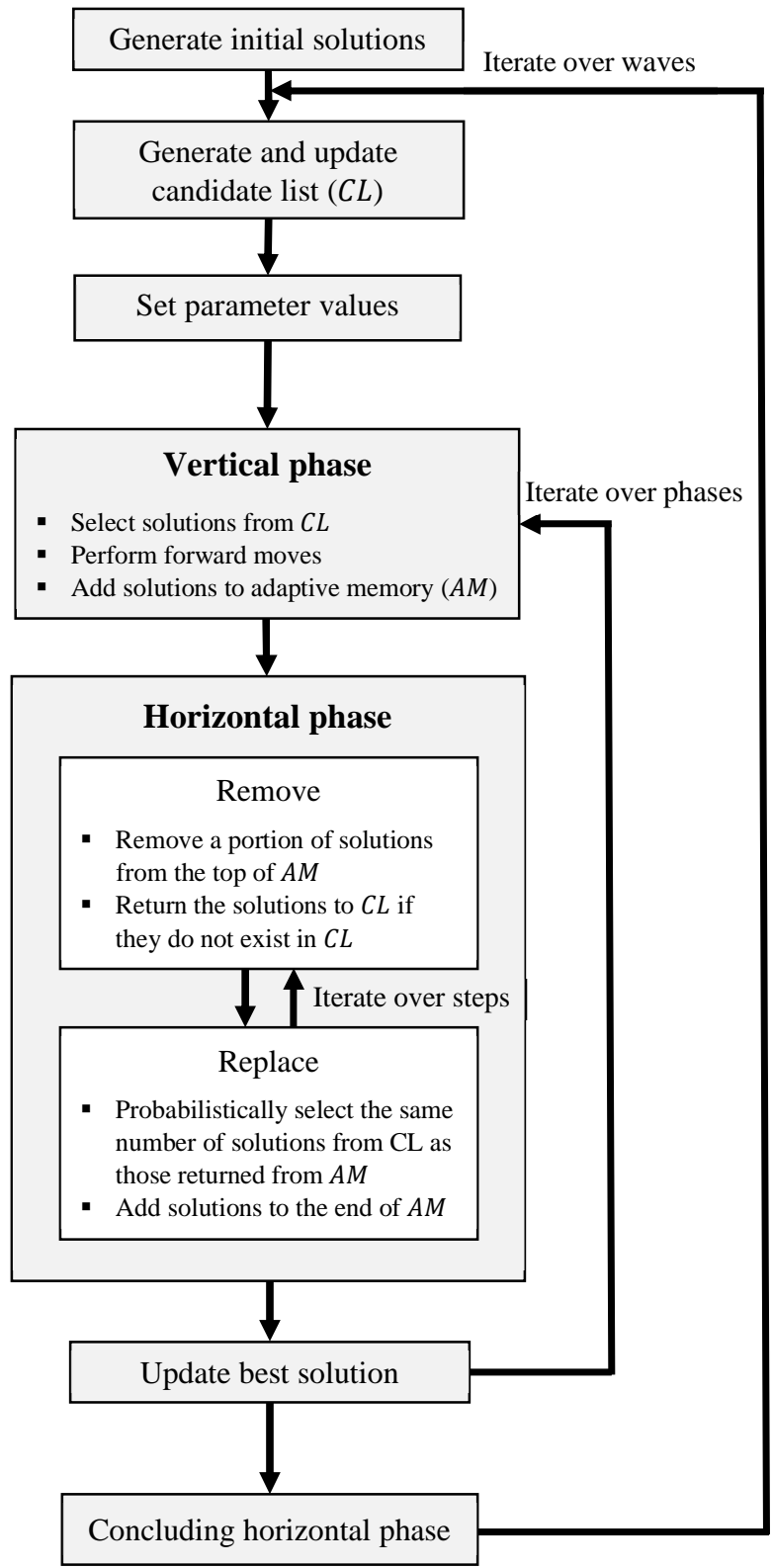

**Fig. 1.** Flow of the M-TAMP algorithm.

The structure of the M-TAMP, as discussed above, is formalized as Algorithm 1. The inputs of the algorithm are: the total number of initial solutions ($I$); the total number of existing crowdsourcee routes ($W$); the number of solutions to be added to *AM* during intervention $q$ ($\mu_q$); the number of horizontal steps during phase $q$ ($h_q$), $\forall q = 1, \ldots, q^*$; and the number of solutions to remove/replace ($d_{i,q}$) during horizontal step $i$ of phase $q$. The output of the algorithm is the best routing solution of crowdsourcees who are assigned with shipping requests ($R_{best}$).

**Algorithm 1.** M-TAMP algorithm

**Input**: $I, W, \mu_q, h_q, d_{i,q}$
**Output**: $R_{best}$

| | | |
|---|---|---|
| 1. | **begin procedure** | |
| 2. | $C_{best} \leftarrow \infty$ | ▷ Initialize DSP total shipping cost |
| 3. | **for** $j = 1, \ldots, I$ | ▷ Consider multiple initial solutions |
| 4. | $R \leftarrow$ Generate an initial solution | ▷ See subsection 4.3.1 |
| 5. | $AM \leftarrow \emptyset$ | ▷ Initialize adaptive memory |
| 6. | **for** $w = 1, \ldots, W$ | ▷ Wave counter |
| 7. | $CL \leftarrow$ Generate a candidate list of solutions | ▷ See subsection 4.3.2 |
| 8. | Set values for $\mu_q, d_{i,q}, h_q$ | ▷ See Appendix A |
| 9. | $CL \leftarrow CL \cup AM$ | ▷ Adding solutions in $AM$ to $CL$ |
| 10. | **for** $q = 1, \ldots, q^*$ | ▷ Phase counter |
| 11. | **if** $|AM| < \mu_q$ **then** | |
| 12. | $Vertical_Phase(\mu_q, AM, CL)$ | |
| 13. | **else** | |
| 14. | $Horizontal_Phase(AM, CL, h_q, d_{i,q}, S_{PA})$ | |
| 15. | **end if** | |
| 16. | $C \leftarrow \min_{R \in AM} C_R$ | ▷ Solution with lowest TSC in $AM$ |
| 17. | **if** $C_{best} > C$ **then** | |
| 18. | $R_{best} \leftarrow R$ | |
| 19. | **end if** | |
| 20. | **end for** | |
| 21. | Perform concluding horizontal phase | |
| 22. | **end for** | |
| 23. | **end for** | |
| 24. | **return** $R_{best}$ | |
| 25. | **end procedure** | |

# 5 Elements of the M-TAMP algorithm

## 5.1 Generating initial solutions

Each time an assignment is performed, M-TAMP involves generating an initial solution first, i.e., perform an initial assignment of each new request to an available crowdsourcee, and then improving the solution. Except for the beginning of the day, the available crowdsourcees for an assignment consist of two types: 1) those who are idle; and 2) those who have assigned requests which are not yet completed, but still have some available time to be assigned more requests. In generating an initial solution, new requests are first added, one at a time, to the end of a crowdsourcee route. Intra-route move is then applied to the added request to obtain the routing sequence with the lowest cost. In this paper, we propose three methods to generate initial solutions based on urgency of a request; crowdsourcee availability of a request; and time availability of a crowdsourcee respectively. Therefore, M-TAMP will start from three initial solutions at each assignment period. The best subsequent solution resulting from the three initial solutions will be chosen for the actual assignment.

A few constraints need to be respected in generating initial solutions. First are feasibility constraints: 1) all requests must be delivered within the guaranteed delivery time; 2) the time to complete a route should not exceed the available time of the crowdsourcee; and 3) the weight carried should not exceed the carrying capacity of the crowdsourcee. In addition, we assume that a crowdsourcee picks up requests only within a distance threshold. Specifically, if a crowdsourcee is idle, the pickup distance of a request is the distance between the pickup location and the current location of the crowdsourcee. If the assigned crowdsourcee is already en route, the pickup distance would be the distance between the end of the pickup location and the existing route.

The first method to generate an initial solution is based on slack time of each request $i$, which is calculated as the difference between the latest possible pickup time $a_i + G_i - T_{p_i,d_i}$ and the current time $t$, where $a_i$ is the time when request $i$ is generated; $G_i$ is the guaranteed delivery time (e.g., 2 hours) of the request; and $T_{p_i,d_i}$ is the crowdsourcee travel time from the pickup to the delivery location on direct route. This calculation, inspired by Mitrović-Minić and Laporte (2004), means that the smaller the slack time, the more urgent a request needs to be delivered. Thus the requests are assigned in order of their urgency to crowdsourcees while satisfying the constraints mentioned above. For a given request, the closest crowdsourcee is considered first. If not possible, then the second closest crowdsourcee, and so on. Once a request is assigned, the time availability of the assigned crowdsourcee will be updated. If a request cannot be assigned to any crowdsourcee, it will be labeled and put aside.

After all requests are checked, unassigned requests that cannot wait any further will be assigned to back-up vehicles which depart from a central depot. A request cannot wait any further if it would be too late to pick up and deliver the request using a backup vehicle at the next assignment period. To ensure speediness, in this paper we assume that each unassigned request will be picked up and delivered by a dedicated back-up vehicle. Unassigned requests that can wait will be left for assignment in the next assignment period.

The second method to generate an initial solution is based on crowdsourcee availability. The idea is to give requests with more limited crowdsourcee availability high priority. Requests are ordered by the number of available crowdsourcees who are within the pickup distance threshold. We look for crowdsourcees to assign for those requests with just one available crowdsourcee, and then requests with two available crowdsourcees, and so on. If two requests have the same number of available crowdsourcees, they are further prioritized based on when they are generated. Similar to the first method, for each request the closest available crowdsourcee is examined first. If not possible to be assigned, then the second closest, etc. Requests that cannot be assigned to any available crowdsourcee will be labeled and put aside. After all the requests are checked, the labeled requests will be similarly treated as in the first initial solution.

The third method considers assigning crowdsourcees instead of requests. Crowdsourcees are assigned in a descending order of available time. The argument for a descending order is that crowdsourcees with greater time availability are more "valuable" considering their possibilities to accommodate future requests. For a given crowdsourcee, we check if the nearest request can be added to meet the feasibility and pickup distance constraints. If added, we update the crowdsourcee's available time and proceed to the crowdsourcee with the highest time availability (which could be the same crowdsourcee if his/her available time is sufficiently large). If adding the request is not possible, we examine the next closest request, and so on. For those requests that end up not assigned, we check again if a request needs to be assigned to a backup vehicle right away or can wait till the next assignment period.

## 5.2 Generating a candidate list of solutions

As mentioned before, once an initial solution is generated for an assignment period, we proceed sequentially with the first wave, the second wave, and so on. The way each wave works is similar and built on the best routing solution of the previous wave. Without loss of generality, the discussion below focuses on the first wave. The first step of a wave is to generate a candidate list of solutions using the initial solution.

Specifically, we remove each request from the route of the wave one at a time. The removed request is added to the end of another route, with intra-route move subsequently performed to generate solutions in the $CL$. To illustrate, suppose we have three routes. A wave is associated with route 1 which has three requests. $3 \times 2 = 6$ solutions are first generated (i.e., removing each of the three requests from route 1 and adding the request to the end of the other two routes). For each of the six solutions, the added request is further moved to an earlier position in the route. This procedure yields a set of alternative solutions. To quantify the attractiveness of a solution $R$, Eq. (1) is used to look at its TSC:

$$C_R = \sum_{m=1}^{M} T_{R,m} + \vartheta T_{R,g} + \tau T_{R,avl} + \rho N_{R,cap} \tag{1}$$

In Eq. (1), TSC is expressed in the amount of time involved. $\sum_{m=1}^{M} T_{R,m}$ denotes the total travel time over all crowdsourcee routes $m = 1, \ldots, M$. $T_{R,g}$ represents the amount of lateness of deliveries with respect to the guaranteed delivery time of the requests. $T_{R,avl}$ indicates the amount of time violation with respect to the available time window of crowdsourcees. $N_{R,cap}$ denotes the total excess weight carried by crowdsourcees. $\vartheta$, $\tau$, and $\rho$ are parameters describing the extent of penalty for the three different types of feasibility violation. $\rho$ also serves to convert the excess weight to time.

Note that the calculation of $C_R$ permits some feasibility constraint to be violated for a solution. We keep some infeasible solutions as they could get improved in M-TAMP. However, if a solution is too inferior, i.e., the cost difference between the solution and the initial solution is too large, it will be discarded. Three pruning techniques are further applied to delete infeasible solutions. First, if the request is too far away from other nodes on the route where the request is moved preventing on-time delivery of the request even by putting the request as the top priority for pickup and delivery, the solution will be deleted. Second, a solution will be pruned if the travel time from any unserved node on the route where the request is moved to the pickup node of the request exceeds the pickup distance threshold, and it is not possible to move another request from a different route to meet the pickup distance threshold while satisfying the feasibility constraints. Third, a solution will be pruned if the crowdsourcee of the route where the request is moved ends his/her available time earlier than the appearance of a request. The remaining solutions after pruning form the candidate list.

## 5.3 Vertical phase

A vertical phase deals with adding solutions to $AM$ using solutions in $CL$ as inputs. The pseudo code is shown in Algorithm 2. The inputs of a vertical phase $q$ are: current $AM$, the number of solutions required to enter $AM$ in the phase ($\mu_q$), and $CL$. The output is an updated $AM$. We iteratively select solutions one from $CL$ at a time (line 3) and perform a forward move (line 5) on each selected solution. The selection of solutions from $CL$ is probabilistic based on cost difference from the initial solution $R_0$ using a Logit-type formula (Eq. (2)). The greater the cost reduction a solution $R$ is compared to $R_0$, the more likely $R$ is selected. Compared to deterministic selection strictly by TSC, probabilistic selection permits diversification: not-so-good solutions in $CL$ get a chance to be selected, explored by a forward move, and enter $AM$.

$$Pr(R) = \frac{\exp\left[a\left(\frac{c_{R_0}-c_R}{c_{R_0}}\right)\right]}{\sum_{R'\in CL}\exp\left[a\left(\frac{c_{R_0}-c_{R'}}{c_{R_0}}\right)\right]} \quad (2)$$

where $a$ is a parameter which can emphasize ($a > 1$) or de-emphasize ($a < 1$) the cost difference.

In Algorithm 2, a forward move (line 4) is performed to improve the selected solution by further exploring its neighborhood. Specifically, after a solution $R_1$ is selected, we remove one more request from the route associated with the current wave and add the removed request to the end of any other routes, thus generating a set of new solutions.

**Algorithm 2.** *Vertical phase* (for phase $q$)
**Input:**   $AMR, \mu_q, CL$
**Output:** Updated $AM$
1.     **begin procedure**
2.        **while** $|AM| < \mu_q$
3.           $R_1 \leftarrow$ Select a solution from $CL$ using Eq. (2)
4.           $R_1'' \leftarrow$ Forward move on $R_1$
5.           Update $AM$
6.           Update $CL$
7.        **end while**
8.        **return** $AM$
9.     **end procedure**

For example, suppose we have three routes and route 1 is associated with the current move. After generating the initial $CL$, only two requests remain in route 1. Then four ($2 \times 2$) new solutions will be generated. A solution will be selected, termed $R_1'$, from the four solutions using Eq. (2). Intra-route moves are then performed on the route to which the request is added. The resulting solution with the least TSC (termed $R_1''$) will be picked. If $TSC_{R_1''} < TSC_{R_1}$, then $R_1''$ enters $AM$ and $R_1$ remains in $CL$. Otherwise, $R_1$ enters $AM$ and also remains in $CL$. $R_1''$ enters $CL$. Solutions in $AM$ are ordered by their times of entry. For a forward move during a wave, if the candidate list is empty or if the associated route is empty we do not perform the forward move.

## 5.4 Horizontal phase

Once the prespecified number of solutions enters $AM$ in a vertical phase, a subsequent horizontal phase is triggered which scrutinizes solutions in $AM$ in the order of their time of entry. Specifically, we check if a solution can be removed from $AM$, returned to $CL$, and replaced with a solution selected from $CL$. This procedure is termed as "removal-replacement". As shown in Algorithm 3, the inputs for a horizontal phase $q$ are: adaptive memory and candidate list, the number of horizontal steps to perform ($h_q$), the number of solutions to remove/replace ($d_{i,q}$) in horizontal step $i$ of phase $q$; and the set of solutions that have entered $AM$ for a sufficient number of times ($S_{PA}$). The output of the algorithm is an updated $AM$ after scrutinizing existing solutions. A horizontal phase is divided into multiple steps $h_q$ (line 2) each performing removal-replacement on a subset of solutions in $AM$. For a given step, if the solution at the top of $AM$ has not appeared in $AM$ enough times (line 6), we return the solution to $CL$ and remove it from $AM$ (lines 7-8). Note that if a solution is already in $CL$, then "return" does not lead to any change to $CL$.

If a solution has appeared in $AM$ enough times, we believe that the solution is attractive enough and move it to the end of $AM$ (lines 10-12). After all solutions are considered in a step, we add the same number of solutions to the end of $AM$ as the number of solutions removed from $AM$ in the step (lines 16-22). The added solutions are probabilistically selected from $CL$ and should not be those already in $S_{PA}$ (lines 17-18). Each time a new solution is added to $AM$, $S_{PA}$ is updated. After all horizontal steps are performed in a horizontal phase, $AM$ is returned for the next vertical phase (line 24).

The removal-replacement procedure updates $AM$ toward better solutions through three aspects. The first aspect is reducing the probability of getting an inferior solution back to $AM$. The "return" operation adds new good solutions generated from forward moves in the previous vertical phase to $CL$. Thus more terms will appear in the denominator of Eq. (2). Consequently, the probability of selecting an inferior solution becomes even smaller. This aspect indeed mimics human brains in that as a person spends time searching for good solutions, he/she will naturally encounter many solutions. When the number of solutions gets larger, it will become more difficult for those not-so-good solutions to come back to the person's memory again.

---

**Algorithm 3.** *Horizontal phase* (for phase $q$)

**Input:** $AM, CL, h_q, d_{i,q}, S_{PA}$

**Output:** Updated $AM$

1.     **begin procedure**
2.         **for** $i = 1, \ldots, h_q$     ▷ $i$ is horizontal step indicator during intervention $q$
3.             $k = 0$     ▷ $k$ records the number of solutions removed from $AM$ in a step
4.             $j = 1$     ▷ $j$ is solution indicator during horizontal step $i$
5.             **while** $j \leq d_{i,q}$     ▷ $d_{i,q}$ is the number solutions at the top of $AM$ list that are to be checked during horizontal step $i$
6.                 **if** $AM[1] \notin S_{PA}$     ▷ Check if the solution on top of the $AM$ list has entered $AM$ enough times
7.                     $CL \leftarrow CL \cup AM[1]$     ▷ If not, add the solution back to $CL$
8.                     $AM \leftarrow AM \setminus AM[1]$     ▷ Remove the solution from the top of $AM$ list
9.                     $k = k + 1$
10.                **else**     ▷ If the solution on top of the $AM$ list has entered $AM$ enough times
11.                     $AM[|AM| + 1] = AM[1]$     ▷ Add the solution to the end of the $AM$ list
12.                     $AM \leftarrow AM \setminus AM[1]$     ▷ Also remove the solution from the top of $AM$ list
13.                **end if**
14.                $j = j + 1$
15.             **end while**
16.             **for** $l = 1, \ldots, k$
17.                **if** $\exists \tilde{R} \in CL$ such that $\tilde{R} \notin S_{PA}$     ▷ Check if there exists one solution in $CL$ that can be added back to $AM$
18.                    $R \leftarrow$ Select a solution from $CL$ using Eq. (2)
19.                    $AM[|AM| + 1] = R$     ▷ Add selected solution to the end of $AM$ list
20.                   Update $S_{PA}$
21.                **end if**
22.             **end for**
23.         **end for**
24.         **return** $AM$
25.     **end procedure**

---

For the second aspect, the use of $S_{PA}$ tabus persistently attractive solutions which are characterized by having appeared in $AM$ enough times, sparing search effort for diversification, i.e., investigating other less explored solutions. This aspect also imitates human memory: in searching for good solutions, solutions that have come across one's head a sufficient number of times will "register" in the person's memory and thus cannot be erased easily. The repeatedly appearing solutions are also likely to be solutions of relatively good quality.

The third aspect relates to dividing a horizontal phase into steps, to increase the chance for not-so-good but maybe promising solutions to be selected from $CL$ into $AM$. As each step only contains a subset of solutions in $AM$, some very good solutions in $AM$, for example those generated from forward moves in the preceding vertical phase, will be deferred to later steps. The "return" operation in the current step then will give greater probability to those not-so-good but promising solutions in $CL$ to be selected into $AM$. This step-wise operation ultimately helps preserve not-so-good but promising solutions in $AM$ at the end of a wave, allowing such solutions to be explored further in future waves. Like the previous two aspects, this aspect is similar to human memory functioning: sometimes a person wants to keep both the best solutions found so far and some currently less good solutions as alternatives in hopes for significant improvement in the future.

## 5.5 Concluding horizontal phase

The purpose of the concluding horizontal phase is to re-examine solutions after the last horizontal phase. The difference between the concluding horizontal phase and a horizontal phase is that in a horizontal phase we remove and replace same number of solutions, whereas in the concluding horizontal phase solutions in $AM$ are replaced until a boundary solution is found (no more selection from $CL$ or forward move possible for a solution selected from $CL$).

# 6 Relocating idle crowdsourcees

The efficiency of the above request-crowdsourcee assignment can be improved with balanced spatial distribution between available crowdsourcees and unassigned requests. However, such balance is unlikely to persist in a crowdshipping system considered in the paper. Recall that the request pickup locations are restaurants, grocery stores, and retail shops which are likely to concentrate in downtown or district centers of a city. In contrast, delivery locations would be residential locations spreading across a city. As a consequence, overtime fewer crowdsourcees will be around the pickup locations of requests. It could also occur that some request pickups are from unpopular locations far away from available crowdsourcees. The imbalance prompts the DSP to use more backup vehicles for guaranteed delivery, resulting in higher TSC.

Proactively relocating idle crowdsourcees in view of unfulfilled and anticipated future demand can mitigate the spatial imbalance. Doing so can also improve crowdsourcee responsiveness. For example, a crowdsourcee relocated in advance to locations close to anticipated requests reduces the travel time for pickup after assignment. However, no efforts have been made in the crowdshipping literature to investigate the issue of crowdsourcee relocation. This section considers relocating idle crowdsourcees between zones which are sub-units of the overall service area. We assume that monetary incentives will be provided for idle crowdsourcees to relocate. Conceptually, an ideal relocation should pursue two objectives that are at odds: 1) minimizing spatial imbalance between request demand and crowdsourcee supply; and 2) minimizing relocation cost. The relocation decision also requires careful consideration of the heterogeneity of the available time among idle crowdsourcees, and how to best use the available time by matching one idle crowdsourcee with multiple unassigned requests while being relocated.

## 6.1 Upper bound for the number of relocatable crowdsourcees

This subsection focuses on finding an upper bound of the number of relocating crowdsourcees. The upper bound is sought assuming that each relocating crowdsourcee would be assigned one request in the zone where he/she is relocated. The upper bound is useful by itself as well as for determining a more cost-effective relocation strategy (subsection 6.2) in which a relocated crowdsourcee can be assigned multiple requests.

We consider that the DSP decides what idle crowdsourcees to relocate right after each assignment, to improve the spatial crowdsourcee-request balance for the next assignment. The relocation also anticipates new arrivals of requests and crowdsourcees. Specifically, after an assignment time $t$, we estimate for each zone $r$ the numbers of idle crowdsourcees ($n_{r,t+\Delta t}$) and unassigned requests ($d_{r,t+\Delta t}$) right before the next assignment at $t + \Delta t$. $n_{r,t+\Delta t}$ is the sum of three terms:

$$n_{r,t+\Delta t} = n'_{r,t} + m_{r,t+\Delta t} + n^{\text{new}}_{r,t+\Delta t} \tag{3}$$

where $n'_{r,t}$ is the number of idle crowdsourcees in zone $r$ right after assignment at $t$. $m_{r,t+\Delta t}$ is the number of crowdsourcees who are en route at $t$ but will finish the assigned delivery trips or relocation trips (which started before $t$) and become available in zone $r$ at $t + \Delta t$. $n^{\text{new}}_{r,t+\Delta t}$ is the expected number of new crowdsourcees arriving in zone $r$ between $t$ and $t + \Delta t$.

$d_{r,t+\Delta t}$ is the sum of two parts:

$$d_{r,t+\Delta t} = d'_{r,t} + d^{\text{new}}_{r,t+\Delta t} \tag{4}$$

where $d'_{r,t}$ is the number of unassigned requests in zone $r$ right after assignment at $t$. $d^{new}_{r,t+\Delta t}$ is the number of new requests arriving in zone $r$ between $t$ and $t + \Delta t$.

Without relocation, the net excess of crowdsourcees ($n^{exc}_{r,t+\Delta t}$) and requests ($d^{exc}_{r,t+\Delta t}$) in zone $r$ right before the next assignment at $t + \Delta t$ can be expressed as Eq. (5)-(6). Obviously, only one of $n^{exc}_{r,t+\Delta t}$ and $d^{exc}_{r,t+\Delta t}$ can be positive (both could be zero).

$$n^{exc}_{r,t+\Delta t} = \max(n_{r,t+\Delta t} - d_{r,t+\Delta t}, 0) \tag{5}$$
$$d^{exc}_{r,t+\Delta t} = \max(d_{r,t+\Delta t} - n_{r,t+\Delta t}, 0) \tag{6}$$

To understand how many crowdsourcees to relocate between two zones, we introduce $\theta_{r,t+\Delta t}$, the fraction of excess requests in zone $r$ among all excess requests in the system at $t + \Delta t$:

$$\theta_{r,t+\Delta t} = \frac{d^{exc}_{r,t+\Delta t}}{\sum_{s \in Z} d^{exc}_{s,t+\Delta t}} \tag{7}$$

It would be desirable to redistribute excess crowdsourcees, which totals $\sum_{r \in Z} n^{exc}_{r,t+\Delta t}$, in proportion to the excess requests across zones. Thus, $\lfloor \theta_{r,t+\Delta t} \sum_{s \in Z} n^{exc}_{s,t+\Delta t} \rfloor$ excess crowdsourcees would be wanted for zone $r$ after such a proportionate relocation is done. The floor operator ensures that the number is integer. The resulting net excess of crowdsourcees in zone $r$ right before the next assignment at $t + \Delta t$, $n^{exc,rloc}_{r,t+\Delta t}$, will be:

$$n^{exc,rloc}_{r,t+\Delta t} = \lfloor \theta_{r,t+\Delta t} \sum_{s \in Z} n^{exc}_{s,t+\Delta t} \rfloor - d^{exc}_{r,t+\Delta t} \tag{8}$$

Note that the above calculation counts the relocated crowdsourcees who may still be on the way to the relocating zone by $t + \Delta t$. Also, for the relocation to occur, it must be that $\sum_{r \in Z} n^{exc}_{r,t+\Delta t} > 0$ and at least one zone has excess requests, i.e., for some zone $r$ $d^{exc}_{r,t+\Delta t} > 0$. Below we present three remarks from investigating the above derivations.

**Remark 1.** If relocation occurs and a zone $r$ has a net excess of crowdsourcees, i.e., $n^{exc}_{r,t+\Delta t} > 0$, then all those crowdsourcees will be relocated out of the zone.

**Proof.** Given that $n^{exc}_{r,t+\Delta t} > 0$, by Eqs. (5)-(6) it must be that $d^{exc}_{r,t+\Delta t} = 0$, which means that $\theta_{r,t+\Delta t} = 0$ by Eq. (7). Plugging $\theta_{r,t+\Delta t} = 0$ and $d^{exc}_{r,t+\Delta t} = 0$ into Eq. (8) leads to $n^{exc,rloc}_{r,t+\Delta t} = 0$. $n^{exc}_{r,t+\Delta t} \geq 0$ and $n_{r,t+\Delta t}^{\text{exc,rloc}} = 0$ together mean that any net excess of crowdsourcees in zone $r$ will be relocated out of the zone. Moreover, no other idle crowdsourcees will be relocated to the zone. ∎

**Remark 2.** If relocation occurs and a zone $r$ has a net excess of requests, i.e., $d_{r,t+\Delta t}^{\text{exc}} > 0$, then this zone will not relocate its idle crowdsourcees to other zones, but only receive idle crowdsourcees (if any) from other zones.

**Proof.** This remark is not difficult to see. Given that $d_{r,t+\Delta t}^{\text{exc}} > 0$, by Eqs. (5)-(6) it must be that $n_{r,t+\Delta t}^{\text{exc}} = 0$. Thus, zone $r$ has zero contribution to $\sum_{s \in Z} n_{s,t+\Delta t}^{\text{exc}}$ number of excess crowdsourcees. On the other hand, zone $r$ will receive $\lfloor \theta_{r,t+\Delta t} \sum_{s \in Z, s \neq r} n_{s,t+\Delta t}^{\text{exc}} \rfloor$ crowdsourcees from other zones. Note that it is possible that no idle crowdsourcee will be relocated to zone $r$, if $\lfloor \theta_{r,t+\Delta t} \sum_{s \in Z, s \neq r} n_{s,t+\Delta t}^{\text{exc}} \rfloor$ is less than one. ∎

**Remark 3.** If relocation occurs, a zone $r$ has a net excess of requests, i.e., $d_{r,t+\Delta t}^{\text{exc}} > 0$, and the total number of excess crowdsourcees is less than the total number of excess requests in the system, i.e., $\sum_{s \in Z} n_{s,t+\Delta t}^{\text{exc}} < \sum_{s \in Z} d_{s,t+\Delta t}^{\text{exc}}$, then the net excess of crowdsourcees after relocation (and before the next assignment), i.e., $n_{r,t+\Delta t}^{\text{exc,rloc}}$, will be non-positive despite crowdsourcee relocation.

**Proof.** Given that $d_{r,t+\Delta t}^{\text{exc}} > 0$ and $\sum_{s \in Z} n_{s,t+\Delta t}^{\text{exc}} < \sum_{s \in Z} d_{s,t+\Delta t}^{\text{exc}}$ for zone $r$, by Eq. (8) the net excess of crowdsourcees after relocation $n_{r,t+\Delta t}^{\text{exc,rloc}} = \left\lfloor \frac{d_{r,t+\Delta t}^{\text{exc}}}{\sum_{s \in Z} d_{s,t+\Delta t}^{\text{exc}}} \sum_{s \in Z} n_{s,t+\Delta t}^{\text{exc}} \right\rfloor - d_{r,t+\Delta t}^{\text{exc}} < \lfloor d_{r,t+\Delta t}^{\text{exc}} \rfloor - d_{r,t+\Delta t}^{\text{exc}} \leq 0$. ∎

With $n_{r,t+\Delta t}$, $d_{r,t+\Delta t}$, $n_{r,t+\Delta t}^{\text{exc}}$, $d_{r,t+\Delta t}^{\text{exc}}$, $\theta_{r,t+\Delta t}$, and $n_{r,t+\Delta t}^{\text{exc,rloc}}$ introduced, the above description still does not tell where an idle crowdsourcees, if to be relocated, should be relocated. To operationalize crowdsourcee relocation, we consider formulating the problem as an integer linear program (ILP) (9)-(12), to be performed right after assignment at $t$.

$$\min \sum_{r \in Z} \sum_{s \in Z} c_{rs} w_{rs} \qquad (9)$$

$$\text{s.t.} \quad n_{r,t+\Delta t} - d_{r,t+\Delta t} + \sum_{\substack{s \in Z \\ s \neq r}} w_{sr} - \sum_{\substack{s \in Z \\ s \neq r}} w_{rs} \geq n_{r,t+\Delta t}^{\text{exc,rloc}} \quad \forall r \in Z \qquad (10)$$

$$\sum_{\substack{s \in Z \\ s \neq r}} w_{rs} \leq n_{r,t}' \quad \forall r \in Z \qquad (11)$$

$$w_{rs} \in \mathcal{N} \cup \{0\} \quad \forall r, s \in Z \tag{12}$$

The intent is to minimize the relocation cost while letting each zone have at least the number of net excess crowdsourcees as in Eq. (8). $w_{rs}$'s are decision variables in the ILP denoting the number of crowdsourcees relocating from zone $r$ to zone $s$. In the objective function (9), $c_{rs}$ is the amount of payment from the DSP to a crowdsourcee relocating from zone $r$ to zone $s$. Given the travel speed of crowdsourcees, $c_{rs}$ is proportional to the distance between $r$ and $s$. Constraint (10) stipulates that the net idle crowdsourcees without relocation $(n_{r,t+\Delta t} - d_{r,t+\Delta t})$ plus relocated crowdsourcees $(\sum_{\substack{s \in Z \\ s \neq r}} w_{sr} - \sum_{\substack{s \in Z \\ s \neq r}} w_{rs})$ should be at least $n_{r,t+\Delta t}^{\text{exc,rloc}}$. Note that if there exists at least one $r \in Z$ that $n_{r,t+\Delta t} - d_{r,t+\Delta t} < n_{r,t+\Delta t}^{\text{exc,rloc}}$, the need for relocation will arise (otherwise, the minimum-cost answer would be no relocation at all). Constraint (11) requires that the number of relocating crowdsourcees from a zone should not exceed the number of idle crowdsourcees in zone $r$ right after assignment at time $t$ ($n'_{r,t}$). Finally, $w_{rs}$'s should be integers (constraint (12)).

Some investigations are performed on ILP (9)-(12). In fact, we find that constraint (11) is redundant. With constraint (11) removed, an integer solution can be obtained by solving its linear program (LP) relaxation. We formalize these as follows.

**Remark 4:** An integer optimal solution of ILP (9)-(12) can be obtained by solving its LP relaxation (9)-(11) with $w_{rs} \geq 0$.

**Proof:** ILP (9)-(12) can be written in an abstract way as: $\min\{cw | Mw \leq b; w \in \mathcal{N} \cup \{0\}\}$, where $M$ is the constraint matrix of dimension $2|Z| \times (|Z|^2 - |Z|)$ and with elements having only values of 0, -1, or 1 as described below (all 1's are light shaded and all -1's are dark shaded). The matrix is a concatenation of two matrices $M_1$ and $M_2$ corresponding to the LHS of constraints (10) and (11) respectively. The first $|Z|$ rows of $M$ is $M_1$. Because we use "$\leq$", the elements in $M_1$ are of the opposite sign of the constraint matrix of constraint (10). For example, for zone 1 (the first row), the first $|Z| - 1$ elements are all 1's which correspond to outflows ($w_{1s}$) whereas the elements corresponding to inflows ($w_{21}, w_{31}, \ldots, w_{|Z|,1}$) are -1's.

The last $|Z|$ rows of $M$, which comprise $M_2$, correspond to constraint (11). In row $|Z| + 1$ of $M$, the first $|Z| - 1$ elements corresponding to outflows from zone 1 ($w_{1s}$'s) are 1's and all other elements are zero. In row $|Z| + 2$ of $M$, the $|Z|$th till the $(2|Z| - 2)$th elements are 1's and all other elements are zero, and so on.

|       | $w_{12}$ | $w_{13}$ | ... | $w_{1,|Z|}$ | $w_{21}$ | $w_{22}$ | ... | $w_{2,|Z|}$ | ... | $w_{i,i-1}$ | $w_{i,i+1}$ | ... | $w_{|Z|,1}$ | $w_{|Z|,2}$ | ... | $w_{|Z|,|Z|-1}$ |
|-------|----|----|----|----|----|----|----|----|----|----|----|----|----|----|----|----|
| $z_1$ | 1 | 1 | ... | 1 | -1 | 0 | ... | 0 | ... | 0 | 0 | ... | -1 | 0 | ... | 0 |
| $z_2$ | -1 | 0 | ... | 0 | 1 | 1 | ... | 1 | ... | 0 | 0 | ... | 0 | -1 | ... | 0 |
| ⋮ | ⋮ | ⋮ | ⋮ | ⋮ | ⋮ | ⋮ | ⋮ | ⋮ | ⋮ | ⋮ | ⋮ | ⋮ | ⋮ | ⋮ | ⋮ | ⋮ |
| $z_{|Z|}$ | 0 | 0 | ... | -1 | 0 | 0 | ... | -1 | ... | 0 | 0 | ... | 1 | 1 | ... | 1 |
| $z_1$ | 1 | 1 | ... | 1 | 0 | 0 | ... | 0 | ... | 0 | 0 | ... | 0 | 0 | ... | 0 |
| $z_2$ | 0 | 0 | ... | 0 | 1 | 1 | ... | 1 | ... | 0 | 0 | ... | 0 | 0 | ... | 0 |
| ⋮ | 0 | 0 | ... | 0 | 0 | 0 | ... | 0 | ... | 0 | 0 | ... | 0 | 0 | ... | 0 |
| $z_{|Z|}$ | 0 | 0 | ... | 0 | 0 | 0 | ... | 0 | ... | 0 | 0 | ... | 1 | 1 | ... | 1 |

It is well known from the theory of integer programming that if the constraint matrix $M$ is totally unimodular (or TUM, when a matrix has all its sub-determinants with values being 0, -1, or 1), then $\min\{cw|Mw \leq b; w \in N \cup \{0\}\}$ can be solved by its LP relaxation $\min\{cw|Mw \leq b; w \geq 0\}$ (Schrijver, 1998). Thus, if we know that $M$ is TUM, then the proof is done. To show this, we first look at $M_1$ and $M_2$ separately. For $M_1$, each column has exactly one 1 and one -1 because each $w_{ij}$ is inflow of only one zone ($j$) and outflow of also only one zone ($i$). (In fact, $M$ is the node-arc incidence matrix of the directed graph which characterize the network of connected zones). Then $M_1$ is known to be TUM (Poincaré, 1900; Schrijver, 1998). For $M_2$, it is also TUM since it is an interval matrix (i.e., a {0,1}-matrix for which each column has its 1's consecutively), which is known to be TUM (Schrijver, 1998). However, concatenation of two TUM matrices is not guaranteed to be always TUM, but we show below this is the case for $M$.

The remaining of the proof leverages the fact that a matrix is TUM if and only if every 2-by-2 submatrix has determinant in 0, -1, or 1 (Fujishige, 1984). Since $M_1$ and $M_2$ are TUM, any 2-by-2 submatrix coming totally from $M_1$ or $M_2$ satisfies this. We only need to check if any 2-by-2 submatrix with one row from $M_1$ and the other row from $M_2$ has determinant in 0, -1, or 1. Note that all elements in $M$ are 0, -1, or 1. The only possibility of violation would be having a 2-by-2 submatrix of $\begin{bmatrix} -1 & 1 \\ 1 & 1 \end{bmatrix}$ or $\begin{bmatrix} 1 & -1 \\ 1 & 1 \end{bmatrix}$. Note that the second row comes from $M_2$, which only has elements of 0 and 1. If the second row has a 0, then any 2-by-2 submatrix involving this row must have a determinant of 0, -1, or 1. Thus, the second row must be $[1 \quad 1]$. In other words, the corresponding columns have the same second-subscript (e.g., $w_{12}$ and $w_{13}$), which indicates the corresponding two flows are out of the same zone to other zones. On the other hand, having $[-1 \quad 1]$ or $[1 \quad -1]$ as the first row from intersecting a row in $M_1$ with two columns only occurs when the corresponding flows are in opposite direction (one inflow and the other outflow), which contradicts. Thus, $M$ does not have any 2-by-2 submatrix of $\begin{bmatrix} -1 & 1 \\ 1 & 1 \end{bmatrix}$ or $\begin{bmatrix} 1 & -1 \\ 1 & 1 \end{bmatrix}$, and thus is TUM. This completes the proof. ∎

Solving ILP (9)-(12) by its LP relaxation is expected to shorten the computation time as LPs are easier to solve than ILPs. As a final note, it is possible that sometimes ILP (9)-(12) (and its LP relaxation) may not have a feasible solution, in which case relocation will not be performed.

## 6.2 Idle crowdsourcee relocation by virtual assignment

ILP (9)-(12) performs relocation based on the numbers of idle crowdsourcees and unassigned requests. However, as a crowdsourcee can pick up and deliver multiple requests after being relocated, it is possible to relocate fewer crowdsourcees. In other words, ILP (9)-(12) gives an upper bound for the number of relocating crowdsourcees. On the other hand, considering each relocated crowdsourcee carrying multiple requests will significantly increase the relocation problem complexity given the combinatorial nature of routing.

To reduce the computational burden for relocation decision, two strategies are considered. First, similar crowdsourcees in terms of location and time availability are grouped into clusters. Second, we construct feasible jobs for each crowdsourcee cluster. Each job consists of a sequence of requests with the minimum cost routing. With the two strategies, the relocation decision reduces to a much smaller-size problem of assigning crowdsourcee clusters to feasible jobs. A key difference here is that the assignment outcome will not be fully executed. Rather, the outcome is only to inform how to relocate idle crowdsourcees. For this reason, such assignment is termed "virtual assignment" in the rest of the paper. Same as subsection 6.1, virtual assignment will be performed right after each real assignment. At time $t$, the virtual assignment is for time $t + \Delta t$ (thus "virtual") anticipating new arrivals of requests and crowdsourcees between $t$ and $t + \Delta t$.

### 6.2.1 Clustering crowdsourcees

We cluster relocatable crowdsourcees using agglomerative hierarchical clustering (AH-clustering) which starts with singleton clusters (i.e., each relocatable crowdsourcees as a cluster) and progressively forms larger clusters by merging existing smaller clusters. The singleton clusters are constructed by selecting $\sum_{\substack{s \in Z \\ s \neq r}} w_{rs}$ idle crowdsourcees that have the most available time in each zone $r$. $w_{rs}$ comes from solving ILP (9)-(12).

The distance between any two clusters $P$ and $Q$ is calculated as:

$$D(P, Q) = D_l(P, Q) + v D_t(P, Q) \tag{13}$$

where $D(P, Q)$ is the distance measure between $P$ and $Q$, which comes from two sources: 1) $D_l(P, Q)$, the location difference between $P$ and $Q$; 2) $D_t(P, Q)$, the time availability difference between $P$ and $Q$. $v$ is the crowdsourcee speed so that the location and time availability differences can be added together. $D_l(P, Q)$ and $D_t(P, Q)$ are calculated as follows:

$$D_l(P, Q) = \frac{1}{|P||Q|} \sum_{p \in P} \sum_{q \in Q} d_l(p, q) \tag{14}$$

$$D_t(P, Q) = \frac{1}{|P||Q|} \sum_{p \in P} \sum_{q \in Q} d_t(p, q) \tag{15}$$

where $d_l(p, q)$ is the Euclidean distance between the locations of crowdsourcees $p$ and $q$. $d_t(p, q)$ is the absolute difference in the amount of available time of crowdsourcees $p$ and $q$.

Given the singleton clusters, we use Eq. (13) to identify the two singleton clusters with the minimum distance. These two clusters are merged to form a new cluster. Then we calculate the distance between the new cluster and all other singleton clusters. We again select two clusters with the minimum distance and form a second new cluster. Each time a new cluster is formed, the two clusters that make up the new cluster are deleted. This iterative process continues until the minimum distance found exceeds some threshold ($\Psi$) which controls crowdsourcee dissimilarity in a cluster. The threshold needs to be carefully chosen. It should not be too large to ensure a job allocated to a cluster in virtual assignment maintains feasibility when the job is assigned to an actual crowdsourcee in the cluster. On the other hand, the threshold should not be small. Otherwise, clustering will forfeit its usefulness as very few clusters will be formed.

After completing clustering, the centroid of each cluster, characterized by the geometric means of the locations and the time availability of crowdsourcees in the cluster, will be used to represent the cluster in virtual assignment, as discussed below.

### 6.2.2 Virtual assignment

Similar to the actual assignment described in Sections 4-5, each time when a virtual assignment is made, we desire to assign as many requests as possible to idle crowdsourcees. As mentioned earlier, the assignment is made between crowdsourcee clusters and jobs that are feasible to each cluster to reduce the computational burden. Each feasible job consists of one or multiple requests with the minimum-cost routing sequence while meeting feasibility constraints (crowdsourcee carrying capacity; each request in the job must be delivered within the guaranteed time; the job needs to be performed within the available time of the crowdsourcee cluster). The available time of a crowdsourcee cluster will be based on the cluster centroid. For a given cluster, the formation of feasible jobs follows Zou and Kafle (2019) and Arslan et al. (2019) as described in Appendix B.

Once feasible jobs are constructed for each cluster, virtual assignment is formulated as a multiple knapsack problem (MKP). A first version can be written as:

$$\max \sum_{i \in I} \sum_{j \in J} \sum_{k \in K} a_{ij} y_{jk} \tag{16}$$

$$\text{s.t.} \quad \sum_{j \in J} y_{jk} \leq N_k \quad \forall k \in K \tag{17}$$

$$\sum_{k \in K} y_{jk} \leq 1 \quad \forall j \in J \tag{18}$$

$$\sum_{j \in J} \sum_{k \in K} a_{ij} y_{jk} \leq 1 \quad \forall i \in I \tag{19}$$

$$y_{jk} \in \{0,1\} \quad \forall j \in J, k \in K \tag{20}$$

where $y_{jk}$ are 0-1 decision variables indicating whether job $j$ is assigned to crowdsourcee cluster $k$. Only $(j,k)$ pairs that associates job $j$ with crowdsourcee cluster $k$ are considered. $a_{ij}$ is a binary parameter indicating whether request $i$ is in job $j$. Thus, the objective is to maximize the number of assigned requests. Constraint (17) stipulates that each cluster should be assigned the number of jobs no more than the number of crowdsourcees in the cluster. This constraint is needed as later each request will be assigned to a crowdsourcee in the cluster. Constraint (18) states that each job is assigned at most one cluster. Similarly, each request should be assigned at most one cluster (Constraint (19)).

The proposed virtual assignment (16)-(20) is a variant of the general $0-1$ knapsack problem (KP) referred to as set-union knapsack problem (SUKP) (Kellerer et al., 2004; Arulselvan, 2014). However, instead of assignment possibility to multiple clusters, SUKP sets target to choose subset of feasible jobs to maximize profit for a single cluster. Note that, each feasible job here corresponds to a subset of unassigned requests. It is already established in the literature, SUKP is strongly NP-hard even for a simple case (each feasible job comprising of two requests (Goldschmidt et al., 1994)). With a view to reducing the size of the problem instance, we use Appendix B for forming feasible jobs beforehand from the requests to reduce the number of possible permutations, allowing us solving the problem in polynomial time. At this point, we provide solution existence criteria for the virtual assignment formulation (16)-(20).

The objective (16) suggests that each unassigned request is treated equally. However, requests are heterogeneous. For example, a request that is remote from relocatable crowdsourcees will requires more travel distance for pickup than a request nearby. To capture the heterogeneity, we consider maximizing the benefit of virtual assignment as realized by using crowdsourcees instead of backup vehicles to pick up and deliver requests. Thus, the benefit can be quantified by comparing TSC without and with crowdsourcees. The TSC without crowdsourcees would be $\sum_{i \in I} c_i$ where $c_i$ is the cost of picking up and delivering request $i$ using a backup vehicle. The TSC with crowdsourcees is $\sum_{j \in J} \sum_{k \in K} c_{jk} y_{jk} + \sum_{i \in I} \sum_{j \in J} \sum_{k \in K} c_i (1 - a_{ij} y_{jk})$ where $c_{jk}$ represents the cost of using a crowdsourcee in cluster $k$ to perform job $j$. $c_{jk}$ is associated with a crowdsourcee traveling from the cluster centroid to the first pickup node in the routing of job $j$ and then traversing all the nodes in the routing. The second term $\sum_{i \in I} \sum_{j \in J} \sum_{k \in K} c_i (1 - a_{ij} y_{jk})$ denotes shipping cost of requests that are assigned to backup vehicles. The benefit of the virtual assignment is thus $\sum_{i \in I} c_i - \sum_{j \in J} \sum_{k \in K} c_{jk} y_{jk} - \sum_{i \in I} \sum_{j \in J} \sum_{k \in K} c_i (1 - a_{ij} y_{jk})$, which is simplified to $\sum_{i \in I} \sum_{j \in J} \sum_{k \in K} (c_i a_{ij} - c_{jk}) y_{jk}$. The virtual assignment problem accounting for request heterogeneity is formulated as:

$$\max \sum_{i \in I} \sum_{j \in J} \sum_{k \in K} (c_i a_{ij} - c_{jk}) y_{jk} \tag{21}$$

s.t.     Constraints (17)-(20)  (22)

Solving MKP (21)-(22) answers whether and if yes to which crowdsourcee cluster a request is assigned. To further assign requests to specific crowdsourcees in a cluster, we randomly pick a request among those assigned to the cluster and assign the crowdsourcee in the cluster with the most available time. Once the virtual assignment is done, the actual action to take is to relocate the crowdsourcees who actually exist to the first pickup node of their assignment job.

As a final note, although en-route crowdsourcees could also be relocated to a place where a future request is anticipated, we do not consider this possibility as in practice it will cause confusion and even frustration by being directed to places with no real requests (only anticipated). This in fact also justifies our dealing of request assignment and crowdsourcee relocation as two separate operations. Related to this point, while M-TAMP could be used to perform virtual assignment, M-TAMP improves solutions by modifying existing routes and routing sequence, which is not relevant to relocation. Neither would MKP be suitable for real assignment as MKP is incapable of exploring inserting a new request in the middle of an existing routing sequence.

# 7 Numerical experiments

The combined M-TAMP for request-crowdsourcee assignment and cluster- and job-based relocation of idle crowdsourcees is numerically investigated in this section. We primarily investigate two problem sizes: a small-size problem with 397 requests and 98 crowdsourcees, and a larger size with 1325 requests and 328 crowdsourcees, each generated over a whole day's crowdshipping operation (8 hours). In subsection 7.1, we first present and discuss the results for the small-size problems in detail, including problem setup, comparison with three popular heuristic methods to gauge the computational performance, benefits of relocation to further reduce the TSC, and results sensitivity to M-TAMP parameter. In subsection 7.2, to keep the paper length we briefly report implementation results for the larger-size problem instances in terms of total shipping cost and computation time, in comparison with the three heuristics. The computational performance is compared with other heuristics, which only consider request-crowdsourcee assignment but not relocation. The M-TAMP algorithm and the cluster- and job-based relocation are coded in C++ on an Intel Core i7 3770 3.40 GHz machine with 12 GB RAM. ILOG CPLEX v12.8 is used to solve ILP and MKP.

## 7.1 Small-size problem: application of M-TAMP with relocation

### 7.1.1 Setup

We consider a squared service area of $6 \times 6$ miles. For relocation, the area is divided into squared zones of size $0.5 \times 0.5$ miles. Thus in total 144 zones. Shipping requests (each pickup and delivery node of a request is randomly generated, independent of each other) and crowdsourcees arrive in each zone following uniform distributions, at a rate of 0.15 and 0.04 per time step respectively (see Appendix D for an illustration of the rate of requests and crowdsourcees generation by zones and the investigated problem instance considered in this Section). We simulate a day (8 hours) of crowdshipping operation, with time step, i.e., the frequency of assignments and relocations, of 10 minutes. Appendix C offers some arguments for the time step choice. Due to the randomness of shipping requests and crowdsourcee generation, simulations are run 10 times. Thus, the results presented below are the averages over the simulations.

We assume crowdsourcees bike to perform pickup and delivery, at 10 mph. The carrying capacity of each crowdsourcee is 10 lbs. Upon arrival in the system, each crowdsourcee has two hours of available time. The pickup distance threshold is 1.67 miles. Request weights are randomly drawn from a normal distribution of 2-7 lbs. The guaranteed delivery time of a request is two hours after the request is generated. When required, backup vehicles will depart from the depot located at the center of the service area. Given the small weight of a request relative to the typical carrying capacity of a backup vehicle (several hundred pounds to several tons), capacity constraints are not considered for backup vehicles. Backup vehicles travel at 20 mph. The DSP pays crowdsourcees at a rate of $7/hour when they are carrying requests. The same rate will be paid to crowdsourcees while being relocated. The operating cost of a backup vehicle follows Kafle et al. (2017) at $68/hour, thus much more expensive than the crowdsourcee pay rate.

Parameters in the M-TAMP algorithm and the relocation model are given values as follows. In M-TAMP, we set $\alpha = 5$; $\eta = 6$; $(\vartheta, \tau, \rho) = (0.25, 0.1, 5)$. The last three parameters relate to the penalty terms in Eq. (1). In idle crowdsourcee relocation, $\Psi$, which is threshold parameter involved in AH-clustering, is set at 0.95 mile.

### 7.1.2 Comparison with other heuristics

To evaluate the performance of the M-TAMP algorithm, TSC is compared with four alternative methods to assign shipping requests to crowdsourcees. The first two methods are: 1) insertion of a request to the end of a crowdsourcee route; 2) insertion plus intra-route move of the inserted requests. The third method is simulated annealing, which mimics material cooling in a heat bath. Simulated annealing is shown to produce reasonably good solutions for large vehicle routing problems faster than other heuristics such as genetic algorithm (Antosiewicz et al., 2013). The fourth method is reactive tabu search (Nanry and Barnes, 2000), a hierarchical heuristic that alternates between neighborhoods to explore search trajectories. Reactive tabu search solves PDPTW using three move neighborhoods that capitalize on the dominance of pairing and precedence constraints. The last two methods also contain insertion and intra-route move.

The first two methods just involve simple operations of requests. As a result, the computation time will be small. However, the solution quality may not be good. Simulated annealing and reactive tabu search are metaheuristics. Thus in principle, the more time given to the search, the better solutions are expected. To make sensible comparisons, we cap the computation time to be the time needed to implement the M-TAMP algorithm. To further assess the benefit of cluster- and job-based relocation, we also present TSCs with only M-TAMP for assignment and with M-TAMP for assignment plus relocation. The resulting TSCs using different solution methods are reported in Fig. 2.

The results clearly show that M-TAMP has superior performance in reducing TSC. As expected, the simplicity of insertion and insertion plus intra-route move produces quite high TSCs. Reactive tabu search appears to be the best among the four alternative methods, but still has significantly higher TSC than M-TAMP. Fig. 2 also demonstrates that relocation is quite effective in reducing TSC, from $1041

to $952. Each time an assignment is performed together with relocation, the assignment takes about 75 seconds. The relocation needs only 1.2 second to complete.

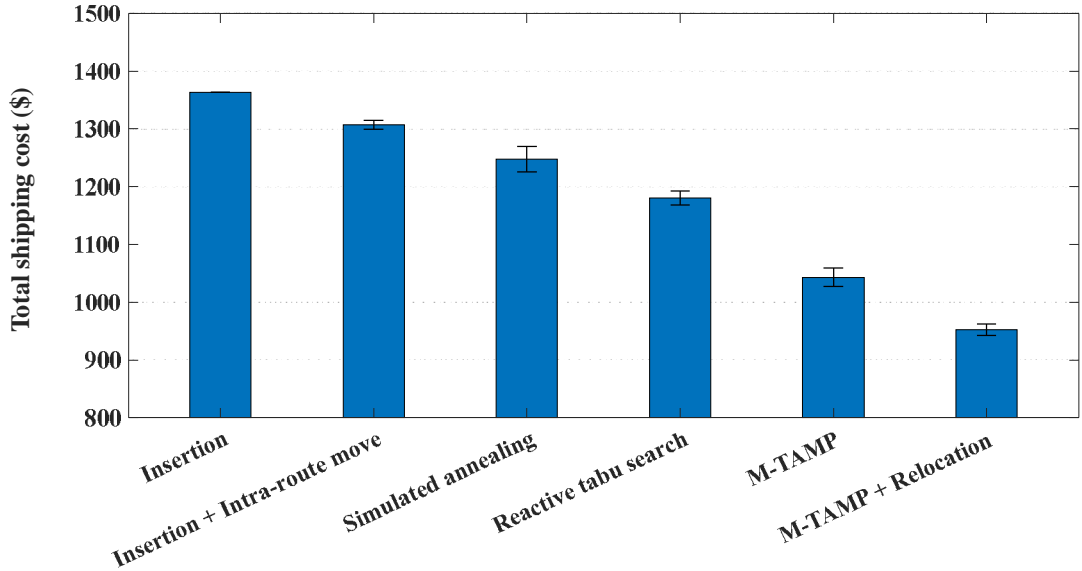

**Fig. 2.** Comparison among different solution methods.

### 7.1.3 Further look into the benefits of relocation

This subsection further evaluates the benefits of relocating idle crowdsourcees from multiple perspectives: TSC, percent of requests fulfilled by crowdsourcees, crowdsourcee availability. Fig. 3 shows the advantage of relocation under a variety of request-to-crowdsourcee ratios. As the ratio increases, the supply of crowdsourcees relative to shipping demand decreases. As a result, the likelihood of having idle crowdsourcees will decrease. The DSP will have less capacity for relocation. Consequently, the difference in TSC with and without relocation becomes smaller.

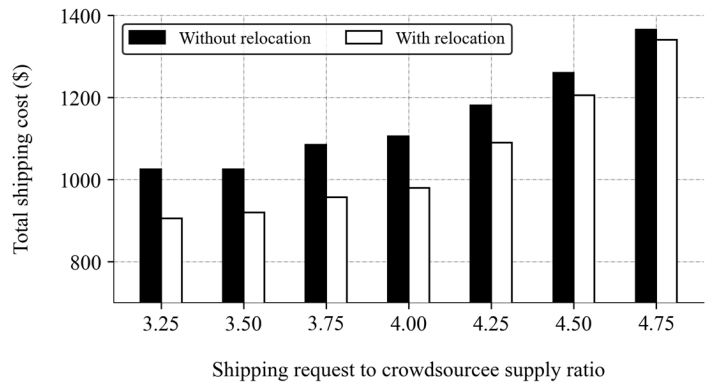

**Fig. 3.** Comparison of total shipping cost with and without relocation under different request-to-crowdsourcee ratios.

Fig. 4 presents the percentage of requests fulfilled by crowdsourcees without and with relocation, as a function of: 1) request-to-crowdsourcee ratio; and 2) request generate rates. In the figure, the more yellow/blue a zone is, the higher/lower the percentage of requests picked up and delivered by crowdsourcees. Focusing on either of the two graphs in Fig. 4, two trends are worth noticing regardless of whether relocation is performed or not. First, keeping the request generation rate constant, an increase in the request-to-crowdsourcee ratio means reduced crowdsourcees in the system. Therefore, fewer requests will be fulfilled by crowdsourcees. Second, holding the request-to-crowdsourcee ratio, an increase in request generate rate implies a proportionate increase in crowdsourcee arrival rate. As the ratios shown in the figure are all above one, in absolute numbers it means greater imbalance between shipping demand and crowdsourcee supply. As a result, we also observe a smaller percentage of requests picked up and delivered by crowdsourcees. Comparing between the two graphs in Fig. 4, we find that relocation, by improving the spatial matching between idle crowdsourcees and demand, increases the percentage of requests fulfilled by crowdsourcees. Given the sheer cost discrepancy between using crowdsourcees and backup vehicles, relocation clearly contributes to reducing TSC. We refer the reader to Appendix E that illustrates the increase of relocation requirements due to increase in the crowdsourcee generation rate (to improve the spatial matching by relocating more idle crowdsourcees).

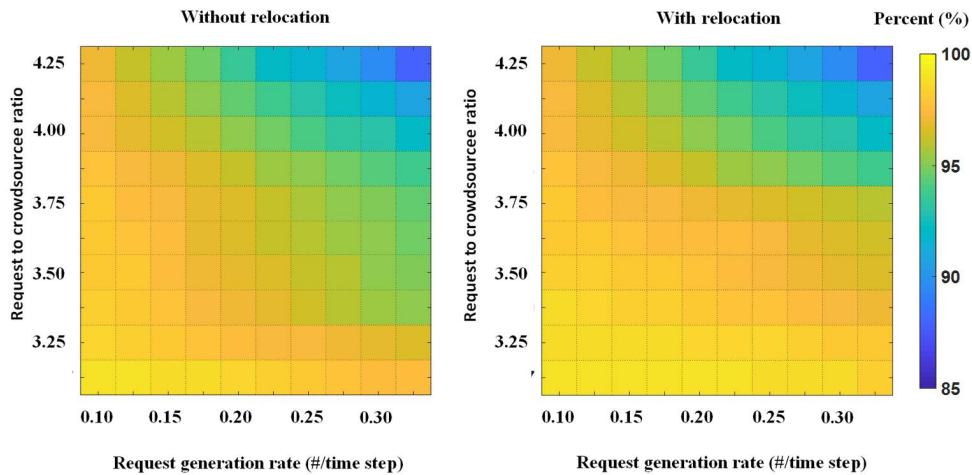

**Fig. 4.** Percentage of requests fulfilled by crowdsourcees without and with relocation.

The benefits of relocating crowdsourcees can be perceived by examining the average number of available crowdsourcees per request, over all zones. Fig. 5 shows this measure at 60, 120, 180, …, 480 mins from the start of the simulated day. Recalling that 10 simulations are performed, boxplots are presented. While there exist natural variations due to the randomness involved in the simulations, the boxplots clearly show that on average greater crowdsourcee availability with than without relocation.

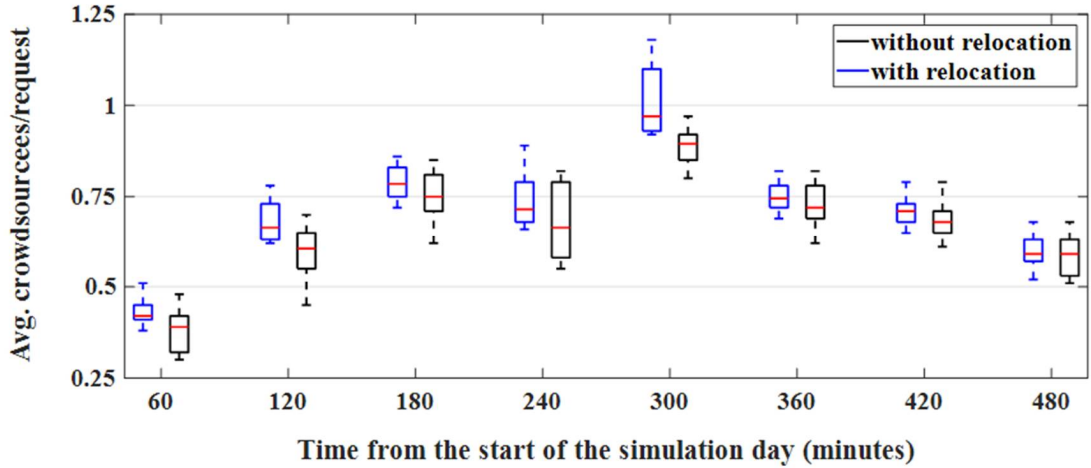

**Fig. 5.** Comparison of the average number of crowdsourcees/request over all zones at different times in a simulation day without and with relocation.

### 7.1.4  Sensitivity of total shipping cost to M-TAMP parameters value

Finally, we investigate the results sensitivity to M-TAMP parameter value: the difference between two successive cardinalities of *AM* values, $\eta$. Fig. 6 presents the results. Each curve in Fig. 6 corresponds to a specific value of the parameter under investigation, for the same randomly generated problem instance considered in Section 7.1.1. While Fig. 6 reports TSC values of one problem instance, we have also experimented with many other randomly generated problem instances and found consistent results. It can be seen that, for all curves in Fig. 6, the chosen value for the parameter produces more reduction in TSC curves than the alternative values. In addition, at any time step, the final TSC using the chosen parameter value is always no worse than using alternative values, which reaffirms our choice of the parameter value.

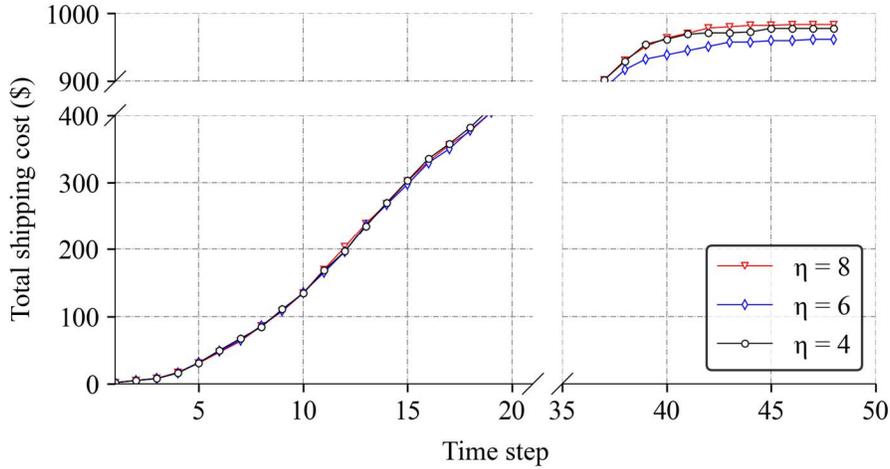

**Fig. 6.** Sensitivity of total shipping cost to parameter value $\eta$

### 7.2 Larger-size problems

#### 7.2.1 Setup

The larger-size problem instance considers problems of assigning 1325 requests to 328 crowdsourcees, which are of comparable size to many dynamic pickup-and-delivery problems investigated in the existing literature (Sayarshad and Gao, 2018; Hyland and Mahmassani, 2018; Ulmer et al., 2021). Apart from a larger number of requests and crowdsourcees, other setups are the same as in the small-size problem setup given in subsection 7.1.1. Shipping requests and crowdsourcees arrive in each zone following a uniform distribution, at a rate of 5.1 and 1.25 per time step respectively. Other parameter values remain the same. With a larger problem size, it is natural to expect a significant increase in computation time for both M-TAMP and other heuristics as given in the following subsection.

#### 7.2.2 Comparison of solutions using M-TAMP and other existing heuristics

Similar to subsection 7.2, this subsection compares the performance of the M-TAMP-based solution approach with the three heuristics (insertion, RTS, and SA). The effectiveness of the M-TAMP-based solution approach is further enforced by considering relocating idle crowdsourcees right after each assignment decision. Fig. 7 shows that M-TAMP with 1-to-n relocation yields the best solution for a whole day simulation consideration. Recall that we find solutions from insertion heuristics are always the worst, despite small computation time. The insertion heuristics yields better result with intra-route move in terms of TSC reduction from $4121 to $3915. While the resulting TSC values from RTS and SA are much comparable to M-TAMP (M-TAMP yields ~4.5% cost reduction over RTS), the computation time is much longer (around 2 minutes, as opposed to ~45 seconds for executing assignment operation at each time step using M-TAMP). By comparing the computation time change between M-TAMP and the same with relocation, it is evident that relocation operation requires very insignificant additional computation effort (on average 3 seconds) while securing ~3.5% further TSC reduction (from $3645 to $3517) compared to M-TAMP-based solution approach. This reduction is less compared to the reduction observed in the small-size problem, which can be attributed to the intuition that with more crowdsourcees in the same service area (thus higher density of crowdsourcees), the need for relocation is reduced. Additionally, it is clear that M-TAMP with relocation is much more scalable than RTS or SA for solving the crowdshipping problems in this paper.

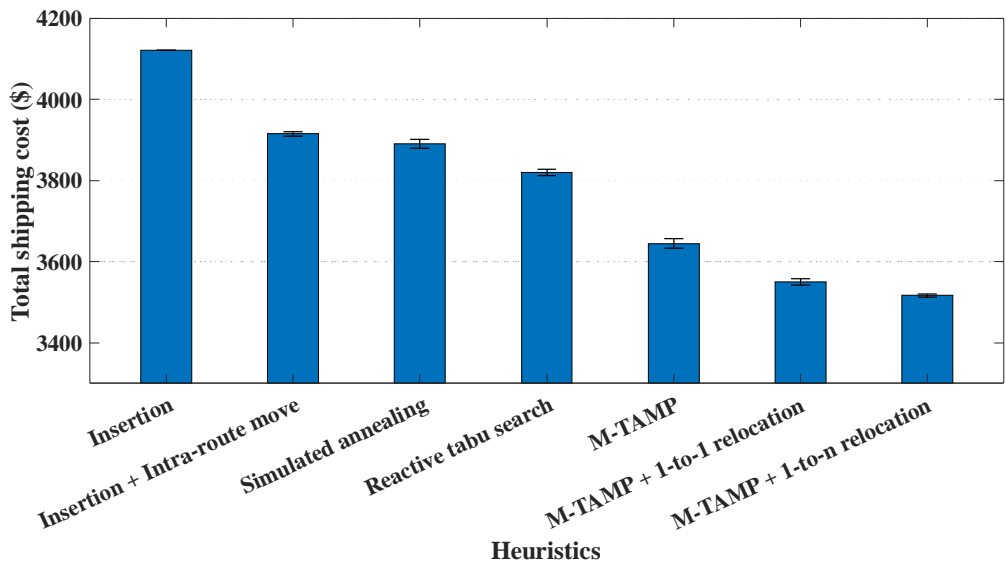

**Fig. 7.** Comparison among different solution methods for the large-size problem.

## 8  Conclusion

This paper investigates two fundamental issues for on-demand crowdshipping: efficient crowdsourcee-request assignment and relocating idle crowdsourcees to further improve assignment efficiency. The on-demand crowdshipping systems considered have spatially distributed request origins and destination and crowdsourcee starting points. In addition, crowdsourcees have limited time availability. Such systems are characteristic of deliveries of food, grocery, and retail in a short time after the order is placed. For the assignment problem, multi-tier adaptive memory programming (M-TAMP) algorithm, a novel population- and memory-based neighborhood search heuristic is proposed. M-TAMP organizes the solution search through waves, phases, and steps, imitating both ocean waves and human memory functioning while seeking the best solution. In view of the short, guaranteed delivery time and the limited time availability of crowdsourcees, three methods are devised to construct initial solutions based on different plausible contemplations in assigning requests to crowdsourcees. All these attributes in M-TAMP contributes to the superiority of the algorithm compared to a number of existing methods.

In addition to using M-TAMP for efficient crowdsourcee-request assignment, this paper also considers proactive relocation of idle crowdsourcees. Two strategies are proposed. The first, more conventional strategy makes relocation decision to rebalance supply-demand based on the numbers of idle crowdsourcees and unassigned requests. The decision is made by solving an ILP. As a crowdsourcee can pick up and deliver multiple requests after being relocated, it is possible to relocate fewer crowdsourcees. In this sense, the first strategy provides an upper bound of the number of relocating crowdsourcees. A cluster- and job-based strategy is further proposed which performs virtual assignment of crowdsourcee clusters to feasible jobs each consisting of one or multiple requests with minimum cost routing. By forming crowdsourcee clusters and feasible jobs, the relocation decision reduces to a small-size, multiple knapsack problem. Numerical experiments show the benefit of relocation to improve the efficiency of system operation, and that the cluster- and job-based relocation can be computed in short time for large problem instances.

The research presented in this paper can be extended in a few directions. First, as argued in the paper, we consider assignment and relocation as two separate operations to avoid possible confusion and frustration of crowdsourcees. Future research can be directed to integrating relocation with assignment while still keeping the routing instructions straightforward to follow. Second, in performing relocation only one period of look-ahead is considered. It would be interesting to explore ways to incorporate greater look-ahead while making relocation decisions. Doing so should recognize the fact that anticipated requests and crowdsourcees will entail greater uncertainty for periods that are farther ahead.

## Acknowledgement

This research is funded by the National Science Foundation under Grant Number 1663411. Earlier versions of this paper were presented at the 2018 INFORMS annual meeting.

## Appendix A: Determining parameter values in M-TAMP

M-TAMP involves several parameters that are related to interventions: the cardinality of $AM$ to maintain, $\mu_q$ during a phase $q$; the number of horizontal steps, $h_q$ required at phase $q$; and the number of solutions to remove/replace ($d_{i,q}$) during a step $i$ of a horizontal phase $q$. At the onset of the M-TAMP algorithm, for the first wave, we generate $CL$ based on section 5.2 and set the value of $\mu_q$ by executing M-TAMP without horizontal phases or examination of both existing and newly added solutions in $AM$. The process initially determines an intermediate measure $\delta$ followed by $q^*$ calculation, thereby finally enabling to calculate $\mu_q$. For setting the value of , we implement the process by continuously adding solutions to $AM$ from $CL$ with forward move (see section 5.3) until the $CL$ is empty or there exists no more forward move for any selected solution (i.e., boundary solution). Similarly, for subsequent waves, we keep on adding solutions to $AM$ and after completion, we can find $\delta = |AM|$.

At this point, we introduce $\eta_q = \mu_q - \mu_{q-1}$ as the difference between two successive cardinalities of $AM$ values. Setting an average difference $\eta \geq 2$, we can determine the total number of phases as $q^* = \left\lfloor \frac{\delta}{\eta} \right\rfloor$, where $\mu_0 = 0$. Note that, $\eta_q$ is bound to be at least 2 because setting $q^* = \delta$ forces the solutions in $AM$ to be re-examined (horizontal phase) after each addition of solution into $AM$ during vertical phase, which is unnecessary. Therefore, during a phase, we can determine the size of $AM$ at which we perform horizontal phase $\mu_q = \frac{(q+1)\delta}{q^*+1}$, $\forall q = 1, \dots, q^*$, where fractional value is rounded to nearest integer. Later, we estimate $d_{i,q}$ during a step $i$ of a horizontal phase $q$. During the first step, we set the value $d_{1,q} < \mu_1 - \mu_0$ (making initially large). With this value, we are ready to estimate $h_q$ for a phase $q$, as $h_q = \left\lceil \frac{|AM|}{d_{1,q}} \right\rceil$. Finally, we calculate the number remove/replacement operation requirement during subsequent steps, $d_{i,q} = \left\lceil \frac{|AM| - \sum_{j=1}^{i-1} d_{j,q}}{h_q - i + 1} \right\rceil$, $\forall i = 2, \dots, h_q$.

## Appendix B: Feasible job formation

Feasible jobs for each crowdsourcee cluster $k$ are formed following a procedure described in Algorithm 4. For each crowdsourcee cluster, the procedure starts by checking the feasibility of all the $N$ unassigned requests (line 5). Specifically, we check if the delivery time, which is the current time plus the time spent for pickup and the travel time between the pickup and delivery nodes, is no later than the latest guaranteed delivery time and the end of the available time of the representative crowdsourcee of the cluster (line 6-8). The check also ensures that the pickup distance does not exceed the threshold (line 8). If these checks are passed for a request, a single-request feasible job is formed.

Starting from the single-request feasible jobs, we progressively add new requests to the jobs (one at a time) and perform intra-route moves to build larger feasible jobs (lines 17-30).

---

**Algorithm 4.** Feasible job formation

| | |
|---|---|
| **Input:** $U, K$ | ▷ $U$ denotes set of unassigned requests |
| **Output:** $J$ | ▷ $J$ denotes set of feasible jobs |

1.  **begin procedure**
2.   **for** $k = 1, \ldots, K$ ▷ Crowdsourcee cluster counter
3.    /* Initial feasible job creation for $k$ */
4.    $L_k \leftarrow$ Location of crowdsourcee type $k$
5.    **for** $j = N + 1, \ldots, 2N$ ▷ Delivery node counter
6.     $t \leftarrow$ current time
7.     $\omega \leftarrow t + t_{L_k, j-N} + t_{j-N, j}$
8.     **if** $\omega \leq \min(a_j + G, e_k)$ and $t + t_{L_k, j-N} \leq t_{thres}$ **then**
9.      Insert $\{j - N, j\}$ at the end
10.     Add the feasible route to $J$
11.    **end if**
12.   **end for**
13.   /* Forming larger jobs for crowdsourcee type $k$ */
14.   **for** $n \in n_\zeta$ ▷ Form larger jobs by repetition
15.    Select a feasible job $\sigma$ from $J$ ▷ Select by probabilistic selection
16.    $L_k \leftarrow$ Set location ▷ At end node on selected job
17.    **for** $j = N + 1, \ldots, 2N$
19.     **if** $j$ is not assigned yet on job $\sigma$ **then**
20.      $t \leftarrow$ Set service end time of $\sigma$ ▷ Ending time of job $\sigma$
21.      $\omega \leftarrow t + t_{L_k, j-N} + t_{j-N, j}$
22.      **if** $\omega \leq \min(a_j + G, e_k)$ & $t + t_{L_k, j-N} \leq t_{thres}$ **then**
23.       $\sigma' \leftarrow \sigma$
24.       Add $\{j - N, j\}$ to the end of job $\sigma'$
25.       Apply intra-route move on $\sigma'$ ▷ Find the cost-minimum sequence
26.       Add route $\sigma'$ to $J$ ▷ New feasible job
27.       **Break**
28.      **end if**
29.     **end if**
30.    **end for**
31.   **end for**
32.  **end for**
33. **end procedure**

## Appendix C: Time step choice in simulation

The DSP requires to set the interval between subsequent assignment attempts considering the rate of shipping request. As we divide the whole service area into zones, we can estimate the probability of receiving at least one new request in zone $r$ during the time interval $\Delta t$.

$$P\{X = k\} = \frac{1}{k!}(\lambda_r \Delta t)^k e^{-\lambda_r \Delta t} \qquad \forall k = 0, 1, \dots, K \tag{C1}$$

$$P\{X > 1\} = P\{X \geq 1\} - P\{X = 1\} = 1 - e^{-\lambda_r \Delta t} - \lambda_r \Delta t * e^{-\lambda_r \Delta t} \tag{C2}$$

$$P\{X > 1\} = 1 - e^{-\lambda_r \Delta t}(1 + \lambda_r \Delta t) \cong 1 - (1 - \lambda_r \Delta t)(1 + \lambda_r \Delta t) = (\lambda_r \Delta t)^2 \tag{C3}$$

$$(\lambda_r \Delta t)^2 \geq 1 - \varepsilon \tag{C4}$$

$$\Delta t \geq \frac{\sqrt{1-\varepsilon}}{\lambda_r} \tag{C5}$$

Therefore, the interval ($\Delta t$) between successive assignment attempts should be short enough $\left(\frac{\sqrt{1-\varepsilon}}{\lambda_r}\right)$ to confirm that the probability of having more than 1 request is at least $(1 - \varepsilon)$. Note that, for the small value of $\lambda t$, we can approximate $e^{-\lambda_r \Delta t} \cong 1 - \lambda_r \Delta t$, as given in Eq. (C3). As the pickup nodes are governed by the distribution of online retail stores (e.g., grocery store, restaurant, etc.) some zones may experience high rate of requests for delivery services than other zones. Assume, the maximum rate of requests is given by $\lambda_r = 0.1$/minute in zone $r$. As a result, providing a 90% guarantee $(1 - \varepsilon = 0.9)$ can be achieved with $\Delta t \geq \frac{\sqrt{0.9}}{0.15} \cong 6.3$ minutes (justifying our choice of 10 minutes).

## Appendix D: Illustration of problem instances used in simulation

We randomly generate the problem instance solved in Section 7.1.1 based on the request and crowdsourcee generation rate as each zone, as given in Fig. D1, in terms of the pickup and delivery locations of the requests, depot location, and origins of the crowdsourcees, as shown in Fig. D2.

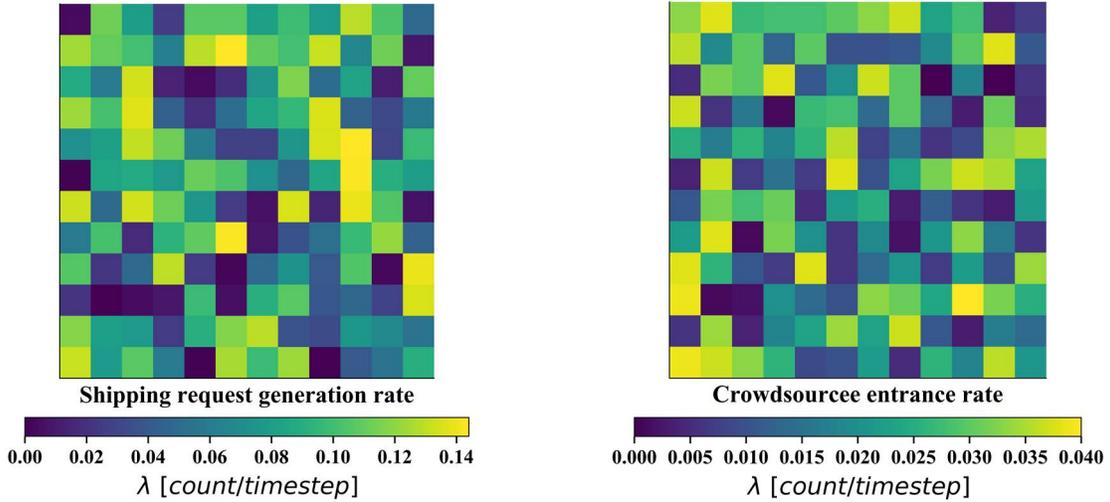

**Fig. D1.** Illustration of the requests and crowdsourcees generation rate at each zone.

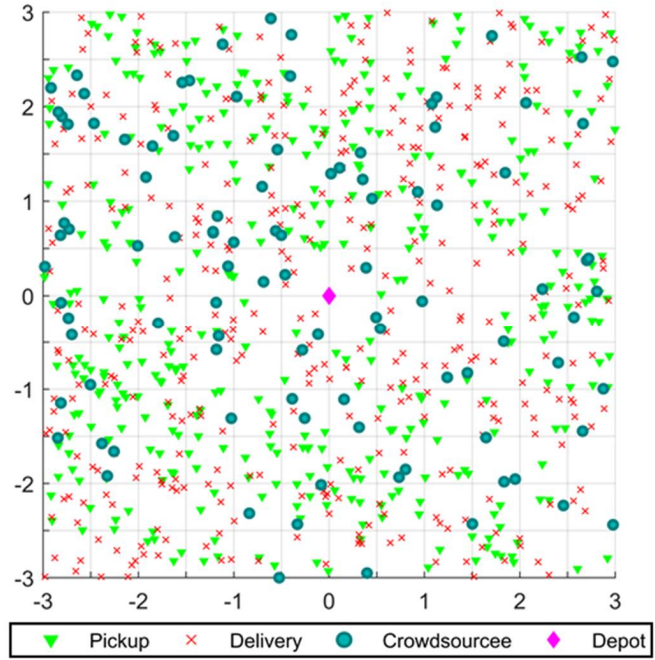

**Fig. D2.** Illustration of the randomly generated problem instance, in terms of the pickup and delivery locations of the requests, depot and origins of the crowdsourcees.

## Appendix E: Illustration of relocation frequency at different crowdsourcee generation rate

Keeping the shipping request generation rate constant, we decrease the crowdsourcee generation rate (0.0400, 0.0376, and 0.0353) and illustrate the frequency of relocation in Fig. E1. We find that not every time step requires relocation right after the assignment. Additionally, as the # of crowdsourcees decreases in the system area, there will be fewer idle crowdsourcees left right after an assignment. Therefore, as the crowdsourcee generation rate decreases, the total number of relocation decisions also decreases. Note that, the top graph of Fig. E1 is the base case.

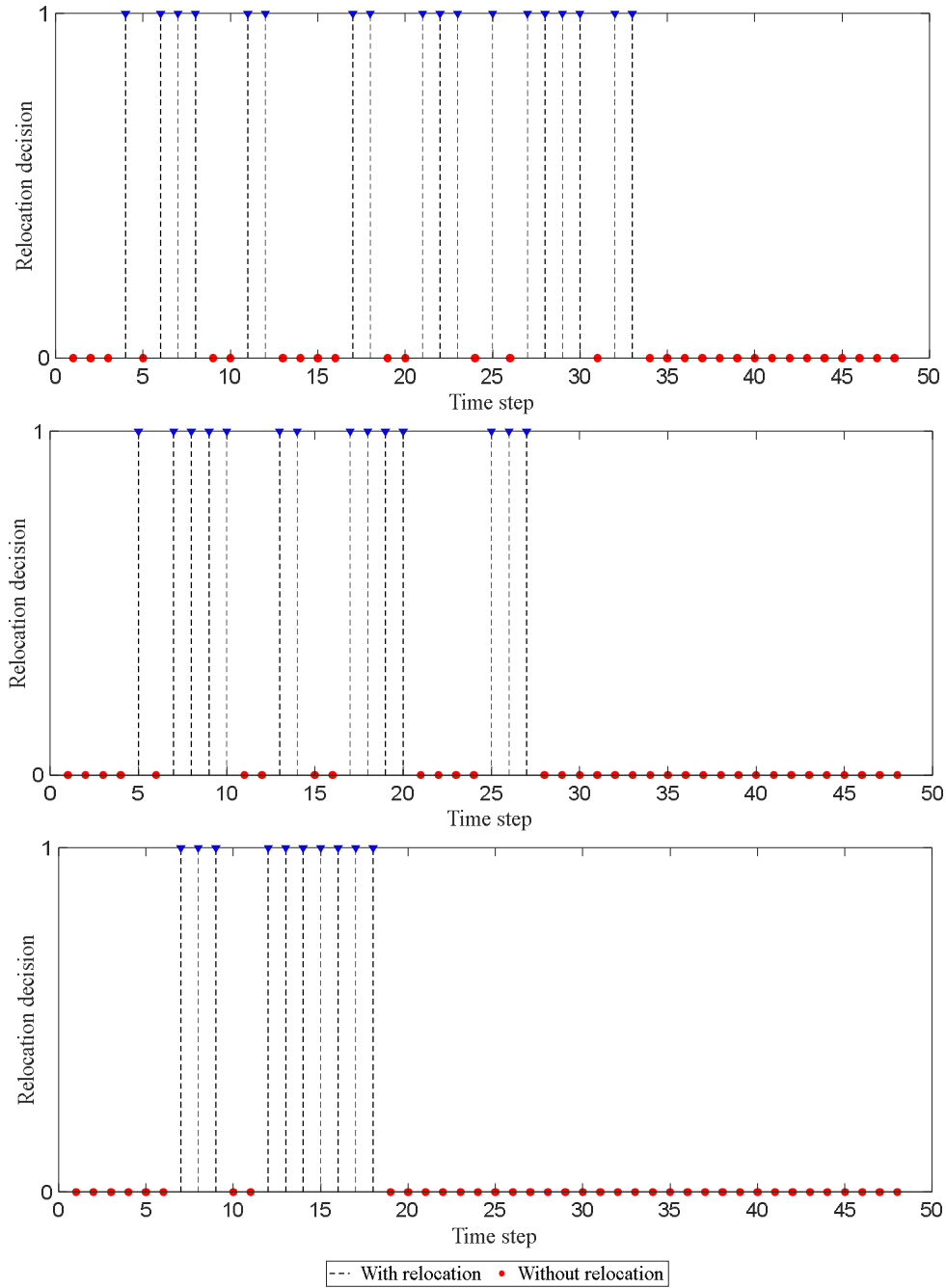

**Fig. E1.** Extent of relocation for different crowdsourcee generation rate.

# References

1. Akeb, H., Moncef, B. and Durand, B., 2018. Building a collaborative solution in dense urban city settings to enhance parcel delivery: An effective crowd model in Paris. *Transportation Research Part E: Logistics and Transportation Review,* 119, pp.223-233.

2. Antosiewicz, M., Koloch, G. and Kamiński, B., 2013. Choice of best possible metaheuristic algorithm for the travelling salesman problem with limited computational time: quality, uncertainty and speed. Journal of Theoretical and Applied Computer Science Vol, 7(1), pp.46-55.

3. Archetti, C., Savelsbergh, M. and Speranza, M.G., 2016. The vehicle routing problem with occasional drivers. *European Journal of Operational Research,* 254(2), pp.472-480.

4. Arslan, A.M., Agatz, N., Kroon, L. and Zuidwijk, R., 2019. Crowdsourced delivery—A dynamic pickup and delivery problem with ad hoc drivers. Transportation Science, 53(1), pp.222-235.

5. Arulselvan, A., 2014. A note on the set union knapsack problem. Discrete Applied Mathematics, 169, pp.214-218.

6. Bent, R. and Van Hentenryck, P., 2004. A two-stage hybrid local search for the vehicle routing problem with time windows. *Transportation Science,* 38(4), pp.515-530.

7. Bent, R. and Van Hentenryck, P., 2007, January. Waiting and Relocation Strategies in Online Stochastic Vehicle Routing. In *IJCAI* (pp. 1816-1821).

8. Berbeglia, G., Cordeau, J.F. and Laporte, G., 2010. Dynamic pickup and delivery problems. *European journal of operational research,* 202(1), pp.8-15.

9. Boyaci, B., Zografos, K.G. and Geroliminis, N., 2017. An integrated optimization-simulation framework for vehicle and personnel relocations of electric carsharing systems with reservations. *Transportation Research Part B: Methodological*, 95, pp.214-237.

10. Dötterl, J., Bruns, R., Dunkel, J. and Ossowski, S., On-Time Delivery in Crowdshipping Systems: An Agent-Based Approach Using Streaming Data.

11. Dayarian, I. and Savelsbergh, M., 2017. Crowdshipping and same-day delivery: employing in-store customers to deliver online orders. *Optimization Online*, pp.07-6142.

12. Devari, A., Nikolaev, A.G. and He, Q., 2017. Crowdsourcing the last mile delivery of online orders by exploiting the social networks of retail store customers. *Transportation Research Part E: Logistics and Transportation Review,* 105, pp.105-122.

13. eMarketer, 2017. E-commerce share of total retail sales in United States from 2013 to 2021. Statista [viewed 30 June 2019]. Available from: https://www.statista.com/statistics/379112/e-commerce-share-of-retail-sales-in-us/

14. Fagnant, D.J. and Kockelman, K.M., 2014. The travel and environmental implications of shared autonomous vehicles, using agent-based model scenarios. *Transportation Research Part C: Emerging Technologies,* 40, pp.1-13.
15. Ferrucci, F., Bock, S. and Gendreau, M., 2013. A pro-active real-time control approach for dynamic vehicle routing problems dealing with the delivery of urgent goods. *European Journal of Operational Research,* 225(1), pp.130-141.
16. Ferrucci, F. and Bock, S., 2014. Real-time control of express pickup and delivery processes in a dynamic environment. *Transportation Research Part B: Methodological,* 63, pp.1-14.
17. Fujishige, S., 1984. A system of linear inequalities with a submodular function on $\{0,\pm1\}$; vectors. Linear algebra and its applications, 63, pp.253-266.
18. Gendreau, M., Guertin, F., Potvin, J.Y. and Séguin, R., 2006. Neighborhood search heuristics for a dynamic vehicle dispatching problem with pick-ups and deliveries. *Transportation Research Part C: Emerging Technologies,* 14(3), pp.157-174.
19. Gdowska, K., Viana, A. and Pedroso, J.P., 2018. Stochastic last-mile delivery with crowdshipping. *Transportation research procedia,* 30, pp.90-100.
20. Ghiani, G., Manni, E., Quaranta, A. and Triki, C., 2009. Anticipatory algorithms for same-day courier dispatching. *Transportation Research Part E: Logistics and Transportation Review,* 45(1), pp.96-106.
21. Glover, F., 1995. Tabu search fundamentals and uses (pp. 1-85). Boulder: Graduate School of Business, University of Colorado.
22. Glover, F. and Laguna, M., 1997a. General purpose heuristics for integer programming—Part I. *Journal of Heuristics*, 2(4), pp.343-358.
23. Glover, F. and Laguna, M., 1997b. General purpose heuristics for integer programming—part II. *Journal of Heuristics*, 3(2), pp.161-179.
24. Glover, F., 2016. Multi-wave algorithms for metaheuristic optimization. *Journal of Heuristics*, 22(3), pp.331-358.
25. Goldschmidt, O., Nehme, D. and Yu, G., 1994. Note: On the set-union knapsack problem. Naval Research Logistics (NRL), 41(6), pp.833-842.
26. Hyland, M. and Mahmassani, H.S., 2018. Dynamic autonomous vehicle fleet operations: Optimization-based strategies to assign AVs to immediate traveler demand requests. Transportation Research Part C: Emerging Technologies, 92, pp.278-297.
27. Kafle, N., Zou, B. and Lin, J., 2017. Design and modeling of a crowdsource-enabled system for urban parcel relay and delivery. *Transportation research part B: methodological,* 99, pp.62-82.

28. Larsen, A., Madsen, O.B. and Solomon, M.M., 2004. The a priori dynamic traveling salesman problem with time windows. *Transportation Science,* 38(4), pp.459-472.
29. Le, T.V., Stathopoulos, A., Van Woensel, T. and Ukkusuri, S.V., 2019. Supply, demand, operations, and management of crowd-shipping services: A review and empirical evidence. *Transportation Research Part C: Emerging Technologies,* 103, pp.83-103.
30. Macrina, G., Pugliese, L.D.P., Guerriero, F. and Laporte, G., 2020. Crowd-shipping with time windows and transshipment nodes. Computers & Operations Research, 113, p.104806.
31. Mitrović-Minić, S. and Laporte, G., 2004. Waiting strategies for the dynamic pickup and delivery problem with time windows. Transportation Research Part B: Methodological, 38(7), pp.635-655.
32. Mitrović-Minić, S., Krishnamurti, R. and Laporte, G., 2004. Double-horizon based heuristics for the dynamic pickup and delivery problem with time windows. *Transportation Research Part B: Methodological,* 38(8), pp.669-685.
33. Nanry, W.P. and Barnes, J.W., 2000. Solving the pickup and delivery problem with time windows using reactive tabu search. *Transportation Research Part B: Methodological,* 34(2), pp.107-121.
34. Nourinejad, M., Zhu, S., Bahrami, S. and Roorda, M.J., 2015. Vehicle relocation and staff rebalancing in one-way carsharing systems. *Transportation Research Part E: Logistics and Transportation Review,* 81, pp.98-113.
35. Pillac, V., Gendreau, M., Guéret, C. and Medaglia, A.L., 2013. A review of dynamic vehicle routing problems. *European Journal of Operational Research,* 225(1), pp.1-11.
36. Poincaré, H., 1900. On the relations between experimental physics and mathematical physics . Gauthier-Villars.
37. Sampaio, A., Savelsbergh, M., Veelenturf, L.P. and Van Woensel, T., 2018. The benefits of transfers in crowdsourced pickup-and-delivery systems. Optimization Online.
38. Sayarshad, H.R. and Gao, H.O., 2018. A scalable non-myopic dynamic dial-a-ride and pricing problem for competitive on-demand mobility systems. Transportation Research Part C: Emerging Technologies, 91, pp.192-208.
39. Sayarshad, H.R. and Chow, J.Y., 2017. Non-myopic relocation of idle mobility-on-demand vehicles as a dynamic location-allocation-queueing problem. *Transportation Research Part E: Logistics and Transportation Review,* 106, pp.60-77.
40. Schrijver, A., 1998. Theory of linear and integer programming. John Wiley & Sons.
41. Smit, S.K. and Eiben, A.E., 2009, May. Comparing parameter tuning methods for evolutionary algorithms. In 2009 IEEE congress on evolutionary computation (pp. 399-406). IEEE.

42. Statista DMO; Statista, 2018. Retail e-commerce sales in the United States from 2017 to 2023 (in million U.S. dollars). Statista [viewed 30 June 2019]. Available from: https://www.statista.com/statistics/272391/us-retail-e-commerce-sales-forecast/
43. Ulmer, M.W., Thomas, B.W., Campbell, A.M. and Woyak, N., 2021. The restaurant meal delivery problem: dynamic pickup and delivery with deadlines and random ready times. Transportation Science, 55(1), pp.75-100.
44. U.S. Census Bureau News, 2019. Quarterly retail e-commerce sales 1st Quarter 2019. US Census [viewed 30 June 2019]. Available from: https://www.census.gov/retail/mrts/www/data/pdf/ec_current.pdf
45. Van Hemert, J.I. and La Poutré, J.A., 2004, September. Dynamic routing problems with fruitful regions: Models and evolutionary computation. In *International Conference on Parallel Problem Solving from Nature* (pp. 692-701). Springer, Berlin, Heidelberg
46. Wang, Y., Zhang, D., Liu, Q., Shen, F. and Lee, L.H., 2016. Towards enhancing the last-mile delivery: An effective crowd-tasking model with scalable solutions. *Transportation Research Part E: Logistics and Transportation Review,* 93, pp.279-293.
47. Yildiz, B. and Savelsbergh, M., 2019. Service and capacity planning in crowd-sourced delivery. *Transportation Research Part C: Emerging Technologies*, 100, pp.177-199.
48. Zhang, R. and Pavone, M., 2016. Control of robotic mobility-on-demand systems: a queueing-theoretical perspective. *The International Journal of Robotics Research*, 35(1-3), pp.186-203.